\documentclass[nofootinbib, showkeys, unsortedaddress, superscriptaddress, notitlepage]{revtex4-1}
\usepackage{graphicx}
\usepackage{amsmath}
\usepackage{amsthm}
\usepackage{dsfont}
\usepackage{mathptmx}
\usepackage{geometry}
\newcommand{\trace}{\mathrm{tr}}
\newcommand{\RR}{\mathds{R}}
\newcommand{\CC}{\mathds{C}}
\newcommand{\ZZ}{\mathds{Z}}
\newcommand{\NN}{\mathds{N}}
\newcommand{\dd}{\mathrm{d}}
\newcommand{\eps}{\varepsilon}
\DeclareMathOperator*{\argmin}{arg\,min}
\theoremstyle{remark}
\newtheorem{remark}{Remark}

\usepackage{hyperref}

\begin{document}

\title{Gel'fand-Yaglom type equations for calculating fluctuations around Instantons in stochastic systems}

\author{T. Schorlepp}
\email{Timo.Schorlepp@rub.de}
\affiliation{Institute for Theoretical Physics I, Ruhr-University Bochum,
             Universit{\"a}tsstrasse 150,
             D-44801 Bochum, Germany}
\author{T. Grafke}
\email{T.Grafke@warwick.ac.uk}
\affiliation{Mathematics Institute, University of Warwick, Coventry CV4 7AL, United Kingdom}
\author{R. Grauer}
\email{grauer@tp1.rub.de}
\affiliation{Institute for Theoretical Physics I, Ruhr-University Bochum,
             Universit{\"a}tsstrasse 150,
             D-44801 Bochum, Germany}

\date{\today}

\begin{abstract}
  In recent years, instanton calculus has successfully been employed to
  estimate tail probabilities of rare events in various stochastic dynamical
  systems. Without further corrections, however, these estimates can
  only capture the exponential scaling. In this paper, we derive a general,
  closed form expression for the leading prefactor contribution of the
  fluctuations around the instanton trajectory for the computation
  of probability density functions of general observables. The key technique is
  applying the Gel'fand-Yaglom recursive evaluation method to the
  suitably discretized Gaussian path integral of the fluctuations,
  in order to obtain matrix
  evolution equations that yield the fluctuation determinant. We
  demonstrate agreement between these predictions and direct sampling
  for examples motivated from turbulence theory.
\end{abstract}

\maketitle

\section{Introduction}
\label{intro}

Quantifying the probability of rare events is extraordinarily difficult:
They are usually too rare to be efficiently observed or sampled, and
at the same time too important to be ignored. A traditional approach
in statistical physics is to phrase the problem as a path integral,
and extract scaling information from a saddle point approximation
(``instanton'' approximation). 

Saddle point techniques have their origin in solid state and quantum
physics~\cite{zittartz-langer:1966,belavin-polyakov-schwartz-etal:1975,thooft:1976},
where also the term ``instanton'' was introduced. The close relation
to large deviation theory was reviewed
in~\cite{grafke-grauer-schaefer:2015}, and its role as
a non-perturbative method to evaluate path integrals
in~\cite{coleman:1979,vainshtein-zakharov-novikov-etal:1982}.

The instanton calculus consists of four steps: First, the instanton is
computed as the classical solution that minimizes the corresponding
action. This step already quantifies the exponential scaling behavior
of the probability density function (PDF) under consideration. Second,
the contribution of fluctuations is taken into account by expanding
the action to second order around the instanton, which yields a
Gaussian path integral. This contribution corresponds to the fluctuation
determinant of the second variation of the instanton action. Depending
on the system at hand, as third and fourth step, one needs to consider
continuous symmetries (zero modes) and the instanton gas,
respectively.

Recently, there has been much activity and progress in numerous
stochastic dynamical systems on the first step, such as the
Kardar-Parisi-Zhang equation~\cite{meerson-katzav-vilenkin:2016},
Ginzburg-Landau equation~\cite{rolland-bouchet-simonnet:2016}, Earth's
climate~\cite{ragone-wouters-bouchet:2017}, biofilm
formation~\cite{grafke-cates-vanden-eijnden:2017} and ocean surface
waves~\cite{dematteis-grafke-onorato-etal:2019}, but, except for the recent
paper~\cite{grafke-schaefer-vanden-eijnden:2021}, progress on the
remaining steps is developed only for specific
applications~\cite{daumont-dombre-gilson:2000,apolinario-moriconi-pereira:2019}. In
this paper, we focus on the second step and develop a general formalism to
compute the contributions of quadratic fluctuations around the instanton 
solution to the path integral for the evaluation of PDFs. We
will present our approach for general finite dimensional Langevin
equations, but with the focus that the developed methods are (in
particular numerically) applicable to large systems of stochastic
ordinary differential equations (SDEs) and finally to stochastic
partial differential equations (SPDEs) relevant in fluid and plasma
turbulence (e.g.\ Burgers, Navier-Stokes and the magnetohydrodynamic
equations). The computation of fluctuations around instantons is the
most important issue in developing a non-perturbative approach to
understanding anomalous scaling in turbulence.

The outline of this paper is as follows: In Section~\ref{sec:1}, we summarize the path
integral formulation of stochastic systems, introduce the instanton solutions,
and clarify the connection with large deviation theory.
Section~\ref{sec:fluctuations} is the central part of this work that contains our
approach to calculating the fluctuation determinant.
The main technical issues that we address in this section are the calculation of the marginal distribution by performing an appropriate integral over all permitted boundary conditions of the fluctuations, and the impact of the discretization of the path integral on the fluctuation matrix and its determinant in particular.
This leads to equations of the Gel'fand-Yaglom type, which can be linearized by a Radon transformation. The resulting simple equations allow the calculation of the fluctuation determinant even for large systems of SDEs and ultimately also SPDEs.
In section~\ref{sec:examples}, we present multiple examples to validate our method and compare its predictions to analytically known results as well as Monte Carlo simulations.
We conclude the paper with a short discussion of our results in section~\ref{sec:conclusion}.

\section{Instantons and Large Deviations}
\label{sec:1}

Consider the stochastic differential equation (SDE)
\begin{equation}
  \label{eq:SDE}
  \dot u + N(u) = \eta\,,\quad u(-T) = u_0\,,
\end{equation}
where the state of the system is described by the vector $u\in\RR^d$
on the time interval $[-T,0]$ (for $T>0$), and the initial value $u_0\in\RR^d$ is deterministic. The (possibly nonlinear)
deterministic term $N : \RR^d \mapsto \RR^d$ will be referred to as
the \emph{drift term}, while stochasticity is introduced via the
$d$-dimensional white-in-time Gaussian noise $\eta$ with covariance
$\chi\in\RR^{d\times d}$ and amplitude $\eps>0$,
\begin{equation}
  \langle \eta_i(t) \eta_j(t')\rangle = \eps\chi_{ij} \delta(t-t')\,.
\end{equation}
Here, $\langle\cdot\rangle$ denotes the ensemble average over noise
realizations. We are interested in the small noise limit $\eps\to0$,
for which the dynamics given by~(\ref{eq:SDE}) are a perturbation of
the deterministic dynamics
\begin{equation}
  \dot u = -N(u)\,,\quad u(-T)=u_0\,,
\end{equation}
which we further assume to have a single fixed point $\bar u$, the
basin of attraction of which covers all of $\RR^d$. Note that we
consider $\chi$ to be independent of $u$, which
corresponds to additive Gaussian noise.

Now, we are interested in (possibly nonlinear) \emph{observables} of
the form $O:\RR^d\mapsto \RR^{d'}$ which represent some quantities of
interest that we wish to measure at the end of our time interval at
$t=0$. For example, we might want to focus on one component of our
final state, or on its average (both cases would have $d'=1$). Due to
the presence of the noise, the observable $O(u(0))$ is a random
variable, and we might want to talk about its PDF $\rho_O$. In
particular, as is common in stochastic field theory, the PDF of the
observable can be written as a path integral. As we will discuss next,
the small noise limit, $\eps\to0$, then corresponds to a
semi-classical limit of this path integral, allowing for an estimate
via saddlepoint approximation and evaluation of the fluctuation
determinant.

\begin{remark}
  In certain applications, one does not actually take the small noise
  limit $\eps \to 0$, but considers a fixed noise strength which may
  correspond e.g. to a given Reynolds number in fluid
  turbulence. Then, in this setup, one usually focuses on the tails of
  the PDF $\rho_O$ at this specific strength of forcing and estimates
  the tail scaling of the PDF using the instanton method. In this
  paper, we will exclusively focus on the small noise limit in order
  to be able to perform a clean expansion in $\eps$. However, we
  remark that for SPDEs with certain scaling invariances, such as the
  Burgers or Navier-Stokes equation, these two limits, i.e.\ small
  noise and large observable amplitude, strictly correspond to each
  other by a suitable rescaling of all variables. For concreteness,
  consider the one-dimensional stochastic Burgers equation in terms of
  physical quantities
  \begin{equation}
    \partial_t u + u \partial_x u - \nu \partial_{xx} u = \eta,  \quad \left< \eta(x,t) \eta(x',t') \right> = \chi(x-x') \delta(t-t')\,,
  \end{equation}
  and take the gradient at one point in space and time
  \begin{equation}
    \label{eq:burgers-gradient-obs}
    O(u(\cdot, t=0)) = \partial_x u(x=0,t=0)\,,
  \end{equation}
  as the observable of interest. Now suppose we want to estimate the PDF
  of this observable at a large observable value of
  \begin{equation}
    |\partial_x u(x=0,t=0)| = a_0\,.
  \end{equation}
  In general, the Burgers equation can be non-dimensionalized by
  introducing a characteristic length scale $x_0$, a characteristic time
  scale $t_0$ and a consistent velocity scale $u_0 = x_0 / t_0$ as well
  as a characteristic strength of the forcing $\chi_0$:
  \begin{equation}
    \tilde{x} = \frac{x}{x_0}, \quad \tilde{t} = \frac{t}{t_0}, \quad \tilde{u} = \frac{u}{u_0}, \quad \tilde{\eta} = \eta \frac{t_0^{1/2}}{\chi_0^{1/2}}\,.
  \end{equation}
  Dropping all tildes, the Burgers equation in terms of
  non-dimensionalized quantities reads
  \begin{equation}
    \partial_t u + u \partial_x u - \mathrm{Re}^{-1} \partial_{xx} u = \frac{\chi_0^{1/2} t_0^{1/2}}{u_0} \eta,  \quad \left< \eta(x,t) \eta(x',t') \right> = \chi(x-x') \delta(t-t')\,,
  \end{equation}
  with the Reynolds number $\mathrm{Re} = u_0 x_0 / \nu$. Adapting the
  time scale to the gradient strength via
  \begin{equation}
    a_0 \overset{!}{=} \frac{u_0}{x_0} = \frac{1}{t_0}\,,
  \end{equation}
  and choosing $x_0 = \sqrt{\nu / a_0}$ then leads to
  \begin{equation}
    \label{eq:burgers-full-nondim}
    \partial_t u + u \partial_x u - \partial_{xx} u = \eta,  \quad \left< \eta(x,t) \eta(x',t') \right> = \eps \chi(x-x') \delta(t-t')\,,
  \end{equation}
  with the noise strength
  \begin{equation}
    \eps = \frac{\chi_0}{\nu} \cdot \frac{1}{a_0^2} \xrightarrow{a_0 \to \infty} 0\,,
  \end{equation}
  as the only dimensionless control parameter, which corresponds
  precisely to the small noise limit that will be treated in the
  remainder of the paper.
\end{remark}

\subsection{Path integral}
\label{sec:JdD}

Formally, the PDF of the observable $O$ can be expressed as
\begin{equation}
  \rho_O(a) = \langle \delta(O(u(0))-a)\rangle\,.
\end{equation}
We can write this as a path integral over all noise realizations
$\eta$ via
\begin{equation}
  \rho_O(a) = \int D\eta \delta(O(u[\eta](0)) - a) \exp\left(-\frac1{2\eps} \int_{-T}^0 \dd t\; (\eta,\chi^{-1}\eta)_d\right)\,,
\end{equation}
where the suitably normalized path density of noise 
realizations is given by the Gaussian
term, and we introduced the $\RR^d$ inner product abbreviated by
$(\cdot,\cdot)_d$. The $\eta$-dependence of the final configuration
$u(0)$ is denoted here explicitly as $u[\eta](0)$.

For convenience, we can perform a change of variables from noise
realizations $\eta$ to field realizations $u$ by inserting the
SDE~(\ref{eq:SDE}) itself,
\begin{equation}
  \label{eq:path-integral}
  \rho_O(a) = \int Du\, J(u)\, \delta(O(u(0))-a) \exp\left(-\frac1{2\eps} \int_{-T}^0 \dd t \; (\dot u + N(u), \chi^{-1}(\dot u + N(u)))_d\right)\,.
\end{equation}
The Jacobian associated with this change of variables, together with a careful treatment of the continuum limit of the stochastic path integral \cite{hunt-ross:1981}, introduces an additional term $J(u)$ in the
prefactor, which will be important later when dealing with the
corrections from fluctuations. For now, we focus on the exponential
term of order $\eps^{-1}$ representing the \emph{action functional} $S[u]$ denoted by
\begin{equation}
  \label{eq:OM-action}
  S[u] = \tfrac12 \int_{-T}^0 \dd t\; \mathcal L(u,\dot u) = \tfrac12 \int_{-T}^0 \dd t \; (\dot u + N(u), \chi^{-1}(\dot u + N(u)))_d\,,
\end{equation}
where we call $\mathcal L(u,\dot u)$ the \emph{Lagrangian}. Written in
this form, the action functional corresponds to the classical
\emph{Onsager-Machlup} action~\cite{machlup-onsager:1953} of the
stochastic process~(\ref{eq:SDE}).\\

For many applications of relevance, the noise covariance $\chi$ is not
necessarily invertible, corresponding to degrees of freedom of the
system that are unforced. This kind of \emph{degenerate forcing}
renders the above formalism unwieldy, as terms involving $\chi^{-1}$
must be treated with care. A standard way to overcome this
complication was proposed by Janssen and
de~Dominicis~\cite{janssen:1976,dominicis:1976} by introducing an
additional \emph{response field} $p$ via
\begin{equation}
  \label{eq:p-def}
  \chi p =  \dot u + N(u)\,.
\end{equation}
With this transformation, the Onsager-Machlup
action~(\ref{eq:OM-action}) is changed into the Janssen-de~Dominicis
action
\begin{equation}
  \label{eq:JdD-action}
  S[u,p] = \int_{-T}^0 \dd t \; \left((p, \dot u + N(u))_d - \tfrac12 (p, \chi p)_d \right) = \tfrac12 \int_{-T}^0 \dd t \; (p,\chi p)_d\,.
\end{equation}
Written like that, the response field can be interpreted as the
\emph{conjugate momentum} of the field variable $u$. Note that we
formally set the action to infinity if $(\dot u + N(u))$ lies in the
kernel of $\chi$. This simply corresponds to the fact that
trajectories $u(t)$ which are impossible to realize with our
degenerate forcing are assigned zero probability. Note also that in the
following derivations, we will treat $\chi$ as invertible, but the
final result will be formulated only in terms of $\chi$ itself. The
derivation remains valid if one were to take the singular limit
carefully.

\subsection{Instantons}
\label{sec:instantons}

The evaluation of the path integral~(\ref{eq:path-integral}) is a
non-trivial task in general. In the small noise limit, $\eps\to0$,
though, we can make use of a saddlepoint approximation, expanding the
action functional around its minimum. In effect, this corresponds to
an infinite dimensional Laplace method to approximate the path
integral. It is noteworthy that this expansion is non-perturbative
with respect to the original SDE~(\ref{eq:SDE}), i.e.~taking every
nonlinearity fully into account. Instead, it corresponds to an
expansion around the most likely pathway $u_I$, the classical
trajectory, also called the \emph{instanton}, for which $\delta
S[u_I]=0$.

More concretely, the instanton is defined as the solution to the
constrained optimization problem
\begin{equation}
  \label{eq:minimization-problem}
  u_I = \argmin_{\substack{u(-T)=u_0\\O(u(0))=a}} S[u]\,.
\end{equation}
A standard way to solve this constrained optimization problem is by
introducing a Lagrange multiplier $\mathcal F\in\RR^{d'}$ to ensure
the constraint $O(u(0))=a$ at the final point $t=0$, to obtain
\begin{equation}
  \tilde S[u] := S[u] + (\mathcal F, O(u(0))-a)_{d'}\,.
\end{equation}
When considering this in the Janssen-de~Dominicis framework, with $\chi
p = \dot u + N(u)$, the first order variation of $\tilde S$ is given
by
\begin{equation}
  \tilde S[u+\delta u] = S[u] + \int_{-T}^0 \dd t \; \left[ (\delta u,
  -\dot p + \nabla N(u)^\top p)_d \right] + (\delta u(0),p(0) + \nabla O(u(0))^\top \mathcal F)_d\,.
\end{equation}
At the trajectory $(u_I,p_I)$ of vanishing first variation we obtain the \emph{instanton
  equations}
\begin{equation}
  \begin{cases}
    \dot u_I + N(u_I) = \chi p_I & u_I(-T) = u_0\\
    \dot p_I - \nabla N(u_I)^\top p_I = 0 & p_I(0) = -\nabla O(u_I(0))^\top \mathcal F_I
  \end{cases}
\end{equation}
The action at the instanton as a function of the observable value $a$,
denoted by $S_I(a)$ is therefore given by
\begin{equation}
  S_I(a) := S[u_I] = \frac1{2} \int_{-T}^0 \dd t \; (p_I, \chi p_I)_d\,.
\end{equation}
At this point, if we are able to find the instanton $(u_I, p_I)$ as
solution of the constrained minimization
problem~(\ref{eq:minimization-problem}), then we have access to the
\emph{exponential scaling} of the PDF of our observable via
\begin{equation}
  \label{eq:prefactor-general}
  \rho_O(a) = Z_\eps(a) e^{- \eps^{-1} S_I(a)}\,,
\end{equation}
for a prefactor component $Z_\eps$ that might still depend on $a$. It is
the goal of the following sections to obtain a set of equations to
compute also, for each $a$ and as $\eps \to 0$, the prefactor $Z_\eps(a)$
to leading order in $\eps$ (the result of which we denote by $Z(a)$) in order to obtain
the full probability density $\rho_O(a)$ with
\begin{equation}
  \label{eq:prefactor-quadratic-approx}
  \rho_O(a) \overset{\eps \to 0}{\sim} Z(a) e^{- \eps^{-1} S_I(a)}\,.
\end{equation}

\begin{remark}
  The above considerations are equivalent to sample path large
  deviation theory, and in particular Freidlin-Wentzell
  theory~\cite{freidlin-wentzell:2012}. In particular, the action
  functional given in equation~(\ref{eq:OM-action}) corresponds
  exactly to the Freidlin-Wentzell rate function for sample paths.
\end{remark}

\section{The contribution of the quadratic fluctuations}
\label{sec:fluctuations}

In this section we derive a general prescription that permits the
computation of the PDF prefactor $Z$
from~(\ref{eq:prefactor-quadratic-approx}) for any Langevin-type
SDE~(\ref{eq:SDE}) with additive noise in the small noise limit $\eps
\to 0$. Concretely, we will show that to leading order (in $\eps$) the
PDF can be approximated by
\begin{align*}
  \rho_O(a) &= (2 \pi \eps)^{-d'/2} \exp \left\{- \frac{1}{2} \int_{-T}^0 \dd t \; \trace \left[ (\nabla \nabla N(u_I(t)), p_I(t))_d Q(t) \right] \right\} \times \\
  & \quad \times
  \left[\det U  \det \left(\nabla O(u_{I}(0)) Q(0) U^{-1} \nabla O(u_{I}(0))^\top \right)\right]^{-1/2} \exp \left\{- \eps^{-1} S_I \right\}\,.
\end{align*}
Here, the prefactor depends on the solution $Q: [-T,0]\mapsto
\RR^{d\times d}$ of a matrix Riccati equation
\begin{equation*}
  \dot{Q} = \chi - Q \nabla N^\top(u_I) - \nabla N(u_I) Q - Q(\nabla \nabla N(u_I), p_I)_d Q, \quad Q(-T) = 0\,,
\end{equation*}
to be evaluated along the instanton trajectory $(u_I,
p_I)$, and $U$ denotes the $d \times d$ matrix
\begin{equation*}
U = 1 + \left(\nabla \nabla O(u_I(0)), {\cal F}_I \right)_{d'} Q(0)\,.
\end{equation*}
Intuitively, the prefactor term quantifies the functional
determinant of the second variation of the action functional, which
can be computed by the evaluation of the Gaussian
path integral representing the fluctuations around the instanton
trajectory. The Riccati equation is then equivalent to an evaluation of the
functional determinant by the Gel'fand-Yaglom method.

It is well known, and has been discussed at length in the 1970s and
1980s in the literature \cite{haken:1976,graham:1977,wissel:1979,hunt-ross:1981,langouche-roekaerts-tirapegui:1982}, that a correct and consistent
discretization of the stochastic path integral is necessary in order
to obtain meaningful results. This is due to the fact that the
fluctuations in the quadratic expansion constitute a Gaussian
stochastic process which is almost surely nondifferentiable, so the
rules of stochastic calculus have to be applied if calculations
involving the fluctuations are done in the continuum limit. While the
SDE~(\ref{eq:SDE}) has additive noise and hence always describes the
same stochastic process, independent of the specific stochastic
calculus in terms of which is interpreted, one has to be more careful
when performing path integral calculations. Consequently, we will
carry out all derivations in a discretized setting and comment
specifically on all instances where the continuum limit is taken.
Prior to this detailed discrete derivation, we briefly discuss some
general aspects of the quadratic expansion in the continuum limit to
give an overview, and also comment on how to evaluate the prefactor
numerically by Monte Carlo methods.

\subsection{Overview in the continuum limit}
\label{subsec:continuum-overview}
In continuum notation~(\ref{eq:path-integral}), the PDF of $O(u(t=0))$ can be written as
\begin{align}
\label{eq:path-integral-w-jacobian}
\rho_O(a) &= \int_{u(-T) = u_0} Du\; \delta(O(u(0))-a) \times \nonumber\\
& \quad \times \exp \left\{ \frac{1}{2}\int_{-T}^0 \dd t \; \trace \left[\nabla N(u) \right] - \frac{1}{2 \eps} \int_{-T}^0 \dd t \left(\dot{u} + N(u), \chi^{-1} \left[\dot{u} + N(u) \right] \right)_d \right\}\,,
\end{align}
where we explicitly included the term of order $\eps^0$ for the
generalized Onsager-Machlup action in the continuum limit.
Once the instanton trajectory $u_I$ given by~(\ref{eq:minimization-problem})
has been found, we insert
\begin{align}
u = u_I + \sqrt{\eps} \delta u
\end{align}
in the path integral in order to expand the action around the instanton, where $\delta u$
will be referred to as the \textit{fluctuations} around the instanton.
In the small noise limit $\eps \to 0$, this expansion then leads to a Gaussian path integral (details can be found in the next section)
\begin{align}
\label{eq:gaussian-pi-continuum}
\rho_O(a) &\overset{\eps \to 0}{\sim} \eps^{-d'/2} \exp \left\{- \eps^{-1} S_I(a)\right\} \int_{\delta u(-T) = 0} D(\delta u)\; \delta(\nabla O(u_I(0))\delta u(0)) \times \nonumber\\
&\times \exp \left\{-\frac{1}{2} \int_{-T}^0 \dd t \left(\delta u, (\nabla \nabla N(u_I), p_I)_d \delta u \right)_d \right\} \times \nonumber\\
&\times \exp \left\{-\frac{1}{2} \left(\delta u(0), (\nabla \nabla O(u_I(0)), {\cal F}_I)_{d'} \delta u(0) \right)_d \right\} \times \nonumber\\
& \times \exp \left\{- \frac{1}{2} \int_{-T}^0 \dd t \left(\delta \dot{u} + \nabla N(u_I) \delta u, \chi^{-1} \left[\delta \dot{u} + \nabla N(u_I) \delta u \right] \right)_d - \trace \left[\nabla N(u_I) \right] \right\}\,,
\end{align}
where $\left(\nabla \nabla N(u_I), p_I \right)_d$ is a shorthand notation for the $d \times d$ matrix
\begin{equation}
  \left[\left(\nabla \nabla N(u_I), p_I \right)_d \right]_{kl} = \left(\partial_k \partial_l N(u_I), p_I\right)_d\,.
\end{equation}
Hence, we see that in a probabilistic sense, the prefactor is given by the expectation
\begin{align}
\label{eq:prefac-expectation}
Z &= \eps^{-d'/2} \bigg< \delta(\nabla O(u_I(0))\delta u(0)) \exp \left\{-\frac{1}{2} \int_{-T}^0 \dd t \left(\delta u, (\nabla \nabla N(u_I), p_I)_d \delta u \right)_d \right\} \times \nonumber\\
&\quad \times \exp \left\{-\frac{1}{2} \left(\delta u(0), (\nabla \nabla O(u_I(0)), {\cal F}_I)_{d'} \delta u(0) \right)_d \right\}\bigg>\,,
\end{align}
where $\delta u$ is a $d$-dimensional Gaussian process on $[-T,0]$ with $\delta u(-T) = 0$ that satisfies the linear SDE
\begin{align}
\label{eq:sde-linear}
\delta \dot{u} + \nabla N(u_I) \delta u = \eta, \quad \left< \eta(t) \eta^\top(t') \right> = \chi \delta(t-t')\,.
\end{align}
Of course this expectation could be evaluated by Monte Carlo simulations of the
SDE~(\ref{eq:sde-linear}), but this suffers from the usual drawbacks of
Monte Carlo methods, and we aim at developing a closed form deterministic
expression for $Z$ instead, that is also cheap to evaluate numerically. However,
in our numerical examples, this possibility to compute the prefactor
provides a good benchmark for our analytical results.

\begin{remark}
  If the drift term $N$ and observable $O$ are polynomials, then the expansion
  of the action around the
  instanton will terminate at a finite order, without considering the small noise
  limit $\eps \to 0$. For concreteness, consider a quadratic drift term and a
  linear observable, which is again
  relevant e.g.\ for the Burgers equation. Then, upon expanding the action
  in~(\ref{eq:path-integral-w-jacobian}), we see that the full prefactor
  $Z_\eps$ for $\eps > 0$, which we define by~(\ref{eq:prefactor-general}),
  will still be given by the expectation
  in~(\ref{eq:prefac-expectation}) (without the $\nabla \nabla O$-term),
  but now $\delta u$ fulfills the nonlinear SDE
  \begin{align}
  \label{eq:sde-ibis}
  \delta \dot{u} + \nabla N(u_I) \delta u + \frac{\sqrt{\eps}}{2} (\delta u, \nabla \nabla N(u_I) \delta u)_d = \eta, \quad \left< \eta(t) \eta^\top(t') 	\right> = \chi \delta(t-t')\,,
  \end{align}
  where we explicitly see the influence of non-Gaussian fluctuations for finite $\eps$.
  Performing Monte Carlo simulations of~(\ref{eq:sde-ibis}) in order to compute the full
  prefactor outside of the small noise limit corresponds to importance sampling of the
  original SDE~(\ref{eq:SDE}) using the instanton. We call this procedure instanton based
  importance sampling (ibis \cite{ebener-margazoglou-friedrich-etal:2019}) and will use it in our numerical
  experiments in order to compare the quadratic and full prefactor.
\end{remark}

Now, our task in this section is to evaluate the Gaussian
path integral~(\ref{eq:gaussian-pi-continuum}). What renders the
problem non-standard are the final time boundary conditions and terms:
$\delta u(t=0)$ is constrained to the
kernel of $\nabla O(u_I(t=0))$. This corresponds to the situation where
possible (infinitesimal) final fluctuations are confined to the
directions in which the value of our observable remains invariant. We
explicitly have to integrate over all boundary conditions of this
subspace of $\RR^d$ (and these boundary conditions also enter the final result
via the $\nabla \nabla O$-term for nonlinear observables).
We will present two alternatives to do so in this paper. The first variant
consists of integrating out the degrees of freedom on the final time
boundary in order to reduce the remaining fluctuation path integral to
Dirichlet 0 boundary conditions. We term this procedure the
\emph{homogenization} of the boundary conditions of the fluctuation
determinant.

The determination of the remaining functional determinant with
Dirichlet 0 boundary conditions of the second variation operator
\begin{align}
\label{eq:second-variation-continuum}
H = (\nabla \nabla N(u_I), p_I)_d + \left[-\frac{\dd}{\dd t} + \nabla N(u_I)^\top \right] \chi^{-1} \left[\frac{\dd}{\dd t} + \nabla N(u_I) \right]
\end{align}
from~(\ref{eq:gaussian-pi-continuum}) is then a standard procedure, and we
explicitly derive Gel'fand-Yaglom like equations for the evaluation of
this determinant. An aspect that has not yet been discussed in detail
in the literature to our best knowledge is the dependence of these
Gel'fand-Yaglom equations on the discretization of the path integral in the
continuum limit. In particular, the functional determinant \emph{does}
indeed depend on the discretization, and it is only the Jacobian
term from the noise-to-field transformation that cancels this discretization
dependence and renders the final result independent of the
discretization choice. For the Gel'fand-Yaglom equation, we therefore
have a freedom of choice of the discretization, as long as we correct this with
the correct corresponding Jacobian. For this reason, we are able to
choose the discretization optimal for computational purposes. We also remark that
there already exists a large body of literature that discusses Gel'fand-Yaglom type
equations in a more functional analytic setting, important references
being~\cite{forman:1987,kirsten-mckane:2003}. 
A useful review is provided by~\cite{dunne:2008}.
In this setup, one usually considers quotients
of functional determinants or regularization procedures such as zeta function
regularization in order to obtain well defined results, and we prefer to work out
the straightforward discretization approach in this paper
(see, however, the related paper~\cite{nickelsen-engel:2011}, 
where the prefactor of the work distribution in one-dimensional Langevin 
systems is calculated directly by adopting the results of 
Kirsten and McKane~\cite{kirsten-mckane:2003} obtained by applying contour 
integration methods to the zeta function of the respective Sturm-Liouville operators).

After following through with this program, we will have obtained a
Gel'fand-Yaglom formula and boundary homogenization procedure that
leads to a closed form representation of the prefactor
contributions. Finally, we will derive the representation
of the PDF prefactor without homogenization of the boundary conditions
that has been stated at the beginning of this section and can more easily
be computed for large system dimensions $d$. In the context of hydrodynamic
shell models, Daumont, Dombre and Gilson \cite{daumont-dombre-gilson:2000} have
derived a related expression for the influence of the quadratic fluctuations on
the PDF prefactor of a one-dimensional observable by path integral calculations,
but their derivation lead to a more complicated procedure, which they also
did not discuss in the continuum limit. Furthermore, Dean, Miao and Podgornik have
derived a similar expression involving algebraic Riccati equations in the 
case of constant coefficients in \cite{dean-miao-podgornik:2019} via the 
Feynman-Kac formula. We adapt their derivation to our problem in 
remark~\ref{rem:feynman-kac}.

\subsection{Quadratic expansion of the discrete action}
\label{subsec:discrete-general-expansion}
The starting point of our derivation is a time-discretized version
of~(\ref{eq:SDE}): For $\alpha \in [0,1]$ and $n \in \NN$, consider
\begin{equation}
  \label{eq:discretized-sde}
  \frac{u_{i+1} - u_i}{\Delta t} + \alpha N\left(u_{i+1}\right) + (1 - \alpha) N\left( u_i\right) = \eta_i, \quad i = 0, \dots, n-1\,,
\end{equation}
with $\Delta t = T / n$. Here, $u_0$ is still chosen deterministically
from the initial condition of~(\ref{eq:SDE}), and $u_1, \dots, u_N$
are $\RR^d$-valued random variables. The discretized white noise
consists of $n$ zero-mean, $\RR^d$-valued Gaussian random variables
$\eta_0, \dots, \eta_{n-1}$ with
\begin{equation}
  \left< \eta_i \eta_j^\top \right> = \frac{\eps}{\Delta t} \chi \delta_{ij}\,.
\end{equation}
The parameter $\alpha$ of the discretization interpolates between the
explicit Euler-Maruyama method for $\alpha = 0$ and the fully implicit
choice $\alpha = 1$. We stress again that any choice of $\alpha$ has
to yield the same continuum limit, and we will use this freedom to
make a computationally optimal choice later on. Now, with this
discretization, the PDF of $O(u(0))$, evaluated at $a \in \RR^{d'}$,
can be written as
\begin{align}
  \label{eq:pdf-discrete-ansatz}
  \rho_O(a) &= \lim_{n \to \infty} \left< \delta(O(u_n) - a) \right>\\
            &= \lim_{n \to \infty} \left(\frac{\Delta t}{2 \pi \eps} \right)^{nd/2} \left( \det \chi \right)^{-n/2} \int_{\RR^d} \left( \prod_{i=0}^{n-1} \dd^d \eta_i \right) \delta(O(u_n) - a) \times \nonumber\\
            &\hspace{4.5cm} \times \exp\left\{- \frac{\Delta t}{2 \eps} \sum_{i=0}^{n-1} \left(\eta_i, \chi^{-1} \eta_i \right)_d\right\}\,.
\end{align}
The next step is to perform a substitution in the integral in order to
be able to integrate over the field $u$ itself. The discrete
transformation rule~(\ref{eq:discretized-sde}) from $\eta_0, \dots,
\eta_{n-1}$ to $u_1, \dots, u_n$ yields the discretization-dependent
Jacobian
\begin{equation}
  \label{eq:jacobian-discrete}
  J_\alpha^{(n)}[u] = \det \left[\left(\frac{\partial \eta_i}{\partial u_j} \right)_{\substack{i=0, \dots, n-1\\j=1, \dots, n}} \right] = \Delta t^{-nd} \det \left[\prod_{i=n-1}^{0} \left(1 + \alpha \Delta t \nabla N(u_{i+1}) \right) \right]\,.
\end{equation}
In the continuum limit $n \to \infty$, $\Delta t \to 0$, this term
asymptotically behaves as
\begin{equation}
  \label{eq:jacobian-limit}
  J_\alpha^{(n)}[u] \overset{n \to \infty}{\sim} \Delta t^{-nd} \exp \left\{ \alpha \int_{-T}^0 \trace \left[ \nabla N(u(t)) \right] \dd t \right\}\,,
\end{equation}
which can easily be seen by noting that the product
in~(\ref{eq:jacobian-discrete}) tends to the solution of the matrix
differential equation
\begin{equation}
  \dot{M}(t) = \alpha \nabla N(u(t)) M(t), \quad M(-T) = 1 \in \RR^{d \times d}\,,
\end{equation}
so its determinant satisfies
\begin{equation}
  \frac{\dd}{\dd t} \det M(t) = \alpha \, \trace \left[ \nabla N(u(t)) \right] \det M(t), \quad \det M(-T) = 1\,,
\end{equation}
by virtue of the general identity
\begin{equation}
 \frac{\dd}{\dd t} \det M(t) = \det M(t) \trace \left[M(t)^{-1} \dot{M}(t) \right]\,.
\end{equation}
Two important observations regarding the
Jacobian~(\ref{eq:jacobian-limit}) are to be made: Firstly, the
exponent is ${\cal O}\left(\eps^0\right)$, so it is of no importance
for the computation of the instanton field itself in the small noise
limit, and secondly, we can consequently naively substitute its
continuum limit everywhere in the following. The
PDF~(\ref{eq:pdf-discrete-ansatz}) after the $\eta \to u$ substitution
thus reads
\begin{align}
 \label{eq:pdf-discrete-u}
\rho_O(a) &= \lim_{n \to \infty} \left(2 \pi \eps \Delta t\right)^{-nd/2} \left( \det \chi \right)^{-n/2} \int_{\RR^d} \left( \prod_{j=1}^{n} \dd^d u_j \right) \delta(O(u_n) - a) \times \nonumber\\
 &\hspace{3cm}\times \exp \left\{ \alpha \int_{-T}^0 \trace \left[ \nabla N(u(t)) \right] \dd t \right\} \exp\left\{- \eps^{-1} S^{(n)}[u]\right\}\,,
\end{align}
where the discretized Onsager-Machlup action is denoted by
\begin{equation}
  S^{(n)}[u] = \frac{\Delta t}{2} \sum_{i=0}^{n-1} \left(\frac{u_{i+1} - u_i}{\Delta t} + \alpha N\left(u_{i+1}\right) + (1 - \alpha) N\left( u_i\right), \chi^{-1} \left[ \dots \right] \right)_d\,,
\end{equation}
where $[\dots]$ is a placeholder for the repetition of the left
argument of the inner product. For this discrete action at order
${\cal O}\left( \eps^{-1} \right)$, we then have to compute the
discrete instanton $u_{I,0}$, $\dots$, $u_{I,n}$ which minimizes the
action under the boundary condition $O(u_{I,n}) = a$, and its
corresponding conjugate momentum $p_{I,0}, \dots, p_{I,n}$. The method of Lagrange multipliers as explained in section~\ref{sec:instantons} can explicitly be incorporated
in the path integral by using the identity
\begin{equation}
\label{eq:delta-fourier}
  \delta(f(x)) = \frac{1}{(2 \pi)^{d'}} \int_{\RR^{d'}} \dd^{d'} k \, \exp\left\{i (k,f(x))_{d'} \right\}\,,
\end{equation}
or, with ${\cal F} = ik\eps$,
\begin{align}
 \label{eq:pdf-discrete-u-wf}
\rho_O(a) &= \lim_{n \to \infty} \left(2 \pi \eps \Delta t\right)^{-nd/2} \left( \det \chi \right)^{-n/2} (2 \pi i \eps)^{-d'}\int_{\RR^{d'}} \dd^{d'} {\cal F}\int_{\RR^d} \left( \prod_{j=1}^{n} \dd^d u_j \right)\times \nonumber\\
 &\times \exp \left\{ \alpha \int_{-T}^0 \trace \left[ \nabla N(u(t)) \right] \dd t \right\} \exp\left\{- \eps^{-1} \left( S^{(n)}[u] + \left({\cal F}, O(u_n) - a \right)_{d'} \right)\right\}\,,
\end{align}

Note that the instanton will typically
be a classical (in the sense of at least $C^2$) minimizer of the
action in the continuum limit, so any numerical scheme or
discretization can in fact be used to determine the instanton without introducing a
systematic error for the following calculations.\\

Once the instanton has been determined for the specific system at hand, we insert the substitution
\begin{equation}
  \label{eq:fluctuation-expansion}
  u_j = u_{I, j} + \sqrt{\eps} \delta u_j, \quad j = 1, \dots, n
\end{equation}
in the integral~(\ref{eq:pdf-discrete-u-wf}), where $\delta u_j$ can be
interpreted as the fluctuations around the instanton at time
$j$. Analogously, we substitute
\begin{equation}
\label{eq:f-expansion}
{\cal F} = {\cal F}_I + \sqrt{\eps}\delta {\cal F}\,,
\end{equation}
where ${\cal F}_I = {\cal F}_I(a)$ is the specific Lagrange multiplier for
the solution of the instanton optimization problem~(\ref{eq:minimization-problem})
with boundary condition $O(u_{I,n}) = a$. Expanding in the small noise
limit $\eps \to 0$ around the
instanton trajectory then yields a Gaussian path integral in the
fluctuations, which we can explicitly evaluate. Concretely, inserting
equation~(\ref{eq:fluctuation-expansion}) and~(\ref{eq:f-expansion}) into the PDF and expanding yields
\begin{align}
 \label{eq:pdf-discrete-u-small-noise}
 \rho_O(a) &= \lim_{n \to \infty} \left(2 \pi \Delta t \right)^{-nd/2} \left( \det \chi\right)^{-n/2} \exp \left\{ \alpha \int_{-T}^0 \trace \left[ \nabla N(u_I(t)) \right] \dd t \right\} \exp \left\{- \eps^{-1} S_I(a) \right\} \times \nonumber\\
 & \times \eps^{-d'/2}\int_{\RR^d} \left( \prod_{j=1}^{n} \dd^d (\delta u_j) \right) \delta(\nabla O(u_{I,n}) \delta u_n) \times \nonumber\\
 &\times \exp\left\{- \delta^2 S^{(n)}[\delta u] - \frac{1}{2}(\delta u_n, (\nabla \nabla O(u_{I,n}), {\cal F}_I)_{d'} \delta u_n)_{d}\right\}\,,
\end{align}
with the second order expansion of the discretized action given by
\begin{align}
\label{eq:second-variation-discrete}
\delta^2 S^{(n)}[\delta u] &= \frac{\Delta t}{2} \bigg( \sum_{i=0}^{n-1} \bigg[ \bigg(\frac{ \delta u_{i+1} - \delta u_i}{\Delta t} + \alpha \nabla N(u_{I,i+1}) \delta u_{i+1} + (1 - \alpha) \nabla N(u_{I,i}) \delta u_i, \nonumber\\
& \quad \chi^{-1} \left[\frac{\delta u_{i+1} - \delta u_i}{\Delta t} + \alpha \nabla N(u_{I,i+1}) \delta u_{i+1} + (1 - \alpha) \nabla N(u_{I,i}) \delta u_i \right] \bigg)_d \nonumber\\
& + \left(\delta u_i, \left(\nabla \nabla N(u_{I,i}), p_{I,i} \right)_d \delta u_{i} \right)_d \bigg] + \alpha \left(\delta u_n, \left(\nabla \nabla N(u_{I,n}), p_{I,n} \right)_d \delta u_{n} \right)_d \bigg)\,,
\end{align}
where we set $\delta u_0 = 0$. The remaining task is to evaluate the Gaussian
integral~(\ref{eq:pdf-discrete-u-small-noise}) efficiently in the
limit $n \to \infty$.

\subsection{Homogenizing the boundary conditions}
\label{subsec:homogeneous-boundaries}

In this section, we reduce the path
integral~(\ref{eq:pdf-discrete-u-small-noise}) with boundary
constraint $\delta u_n \in \ker \nabla O(u_{I,n})$ to an equivalent
problem with Dirichlet 0 boundary conditions $\delta u_n = 0$. In the
following, we will assume that the linear map $\nabla O(u_{I,n}):
\RR^{d} \to \RR^{d'}$ has full rank $d' \leq d$ for our notational
convenience. We then introduce an orthonormal basis $\left\{ \delta
u_n^{(1)}, \dots, \delta u_n^{(d-d')}\right\}$ of the linear subspace
\begin{equation}
  \ker \nabla O(u_{I,n}) \subset \RR^d\,
\end{equation}
and extend this basis to an orthonormal basis $\left\{ \delta u_n^{(1)}, \dots, \delta u_n^{d-d'}, v^{(1)}, \dots, v^{(d')} \right\}$ of $\RR^d$. Writing $\delta u_n$ in terms of these basis vectors as
\begin{equation}
 \delta u_n = \sum_{i=1}^{d-d'} \beta_i \delta u_n^{(i)} + \sum_{j=1}^{d'} \gamma_j v^{d'}\,,
\end{equation}
the fluctuations in the $v_i$-directions are irrelevant for the
boundary integral over $\delta u_n$
in~(\ref{eq:pdf-discrete-u-small-noise}). Therefore, after changing to
this basis, we can drop the subspace constraint
in~(\ref{eq:pdf-discrete-u-small-noise}) and only integrate over the
remaining $d-d'$-dimensional relevant subspace, which yields
\begin{align}
 \label{eq:bvpint-beta}
  &\int_{\RR^d} \dd^d(\delta u_n) \; \delta(\nabla O(u_{I,n}) \delta u_n) \times \nonumber\\
  &\quad \times \exp \left\{- \delta^2 S^{(n)}(\delta u_0 = 0, \delta u_1, \dots, \delta u_{n-1}, \delta u_n) - \frac{1}{2}(\delta u_n, (\nabla \nabla O(u_{I,n}), {\cal F}_I)_{d'} \delta u_n)_{d} \right\} \nonumber \\
  &=\left[ \det \left( \nabla O(u_{I,n}) \nabla O(u_{I,n})^\top \right) \right]^{-1/2} \int_{\RR^{d-d'}} \dd^{d-d'} \beta \times \nonumber\\
  &\quad \times \exp \left\{- \delta^2 S^{(n)} \left(0, \delta u_1, \dots, \delta u_{n-1}, \sum_{i=1}^{d-d'} \beta_i \delta u_n^{(i)} \right) \right\} \times \nonumber\\
  &\quad \times \exp \left\{ - \frac{1}{2} \sum_{i,j = 1}^{d-d'} \beta_i \beta_j (\delta u_n^{(i)}, (\nabla \nabla O(u_{I,n}), {\cal F}_I)_{d'} \delta u_n^{(j)})_{d} \right\}\,.
\end{align}
Now, by interchanging the order of integration
in~(\ref{eq:pdf-discrete-u-small-noise}), the integral can be
interpreted in the sense that for each individual, \textit{fixed}
boundary condition
\begin{equation}
  \delta u_n^* = \sum_{i=1}^{d-d'} \beta_i^* \delta u_n^{(i)}\,,
\end{equation}
the remaining $(d \cdot (n-1))$- dimensional integral over the integrand
\begin{equation}
  \exp \left\{- \delta^2 S^{(n)} \left( \delta u_0 = 0, \delta u_1, \dots, \delta u_{n-1},  \delta u_n^* \right) \right\}\,,
\end{equation}
has to be carried out for this particular boundary condition. What we
propose to do is to perform, for each fixed boundary condition $\delta
u_n^*$, a shift in the other integration variables:
\begin{equation}
 \delta u_i = \delta u_i^* + \delta \tilde{u}_i, \quad u = 1, \dots, n-1\,,
\end{equation}
such that integration is then performed over $(\delta \tilde{u}_i)_{1 \leq i \leq n-1}$ instead and we demand that
\begin{align}
 \label{eq:quadratic-instanton}
 &\delta^2S^{(n)}(\delta u_0 = 0, \delta u_1, \dots, \delta u_{n-1}, \delta u_n^*) \nonumber\\
&\overset{!}{=} \delta^2S^{(n)}(0, \delta u_1^*, \dots, \delta u_{n-1}^*, \delta u_n^*) + \delta^2S^{(n)}(0, \delta \tilde{u}_1, \dots, \delta \tilde{u}_{n-1}, 0)\,.
\end{align}
Effectively, this corresponds to the condition that the first order
variation (with fixed end points) of the quadratic action should
vanish at the $\delta u_i^*$-trajectory, so we compute additional
instantons for each of the given boundary condition $\delta u_n^*$. If
this cannot be solved analytically, it is of course hopeless to do
this numerically for every single boundary condition, but, since the
action is quadratic at this stage, it suffices to determine these
$\delta u_i^*$ trajectories once for each of the basis vectors $\delta
u^{(1)}, \dots, \delta u^{(d-d')}$. In the continuum limit, which can
again be taken naively for these additional, differentiable
instantons, the condition~(\ref{eq:quadratic-instanton}) can be
written in terms of a linear boundary value problem (BVP)
\begin{align}
\label{eq:bvp}
\begin{cases} \frac{\dd}{\dd t} \left( \begin{array}{c}
\delta u\\
\delta p
\end{array} \right) = \left( \begin{array}{cc}
-\nabla N(u_I) & \chi\\
( \nabla \nabla N(u_I),p_I )_d & \nabla N(u_I)^\top
\end{array} \right)  \left( \begin{array}{c}
\delta u\\
\delta p
\end{array} \right)\,, \\[.4cm]
\delta u(-T) = 0, \; \delta u(0) = \delta u^{(i)}_n\,,
\end{cases}
\end{align}
for each of the $(d-d')$ basis vectors $\delta u^{(i)}_n \in \ker
\nabla O(u_I(0)) \subset \RR^d$. Here, analogously
to~(\ref{eq:p-def}), we introduced the \emph{adjoint fluctuations}
\begin{equation}
 \label{eq:dp-def}
  \chi \delta p = \delta \dot{u} + \nabla N(u_I) \delta u\,,
\end{equation}
in order to reduce the differential equation of the BVP~(\ref{eq:bvp})
to first order. Note that the differential equation~(\ref{eq:bvp}) is
equivalent to $H \delta u = 0$ where $H$ is the second variation
operator~(\ref{eq:second-variation-continuum}) in the continuum
limit. Denoting the discrete solution of the BVP~(\ref{eq:bvp}) for
the basis vector $\delta u_n^{(i)}$ as boundary condition by
\begin{equation}
\left( \delta u_1^{(i)}, \dots, \delta u_{n-1}^{(i)}, \delta u_n^{(i)} \right)\,,
\end{equation}
we can then use the linearity of the corresponding BVPs to expand
\begin{align}
&\delta^2 S^{(n)} \left( \delta u_0 = 0, \delta u_1, \dots, \delta u_{n-1},  \sum_{i=1}^{d-d'}\beta_i \delta u^{(i)}_n \right) \nonumber \\
&= \delta^2 S^{(n)} \left( \delta u_0 = 0, \sum_{i=1}^{d-d'}\beta_i \delta u^{(i)}_1, \dots, \sum_{i=1}^{d-d'}\beta_i \delta u^{(i)}_{n-1}, \sum_{i=1}^{d-d'}\beta_i \delta u^{(i)}_n \right) \nonumber\\
& \quad + \delta^2 S^{(n)}(\delta u_0 = 0, \delta \tilde{u}_1, \dots, \delta \tilde{u}_{n-1}, \delta u_n = 0)\,,
\end{align}
for any given boundary condition, which completely separates the inner
integral over $\delta \tilde{u}_1$, $\dots$, $\delta \tilde{u}_{n-1}$ with
Dirichlet 0 boundary conditions as desired. The remaining integral
over all boundary conditions~(\ref{eq:bvpint-beta}) is a $d-d'$
dimensional Gaussian integral in $\beta$ and can easily be evaluated
in the continuum limit by noticing that for differentiable curves, the
continuum limit of the second variation of the action, written as a
quadratic form, is simply
\begin{equation}
 \delta^2 S[u,w] = \frac{1}{2} \int_{-T}^0 \dd t \left(\dot{u} + \nabla N(u_I) u, \chi^{-1} \left[\dot{w} + \nabla N(u_I) w \right] \right)_d + (u, (\nabla \nabla N(u_I), p_I)_d w)_d\,,
\end{equation}
so from~(\ref{eq:bvp}), we obtain
\begin{align}
\delta^2 S\left[\delta u^{(i)},\delta u^{(j)}\right] &= \frac{1}{2} \int_{-T}^0 \dd t \left( \delta p^{(j)}, \delta \dot{u}^{(i)} + \nabla N(u_I) \delta u^{(i)} \right)_d \nonumber\\
& \quad + \left( \delta u^{(i)}, \left( \nabla \nabla N(u_I),p_I \right)_d \delta u^{(j)} \right)_d = \frac{1}{2} \left( \delta u^{(i)}(0), \delta p^{(j)}(0) \right)_d\,,
\end{align}
for any two solutions $\delta u^{(i)}$, $\delta u^{(j)}$, $1 \leq i,j
\leq d-d'$ of the BVP~(\ref{eq:bvp}). Here, $\delta p^{(j)}$ of course
denotes the adjoint fluctuation~(\ref{eq:dp-def}) for the solution
$\delta u^{(j)}$. Therefore, the $\beta$-integral
in~(\ref{eq:bvpint-beta}) can be evaluated to yield
\begin{align}
&\int_{\RR^{d-d'}} \dd^{d-d'} \beta \, \exp \left\{- \delta^2 S^{(n)} \left( \delta u_0 = 0, \delta u_1, \dots, \delta u_{n-1}, \sum_{i=1}^{d-d'} \beta_i \delta u_n^{(i)} \right) \right\} \times \nonumber\\
& \quad \times\exp \left\{-\frac{1}{2} \sum_{i,j = 1}^{d-d'} \beta_i \beta_j (\delta u_n^{(i)}, (\nabla \nabla O(u_{I,n}), {\cal F}_I)_{d'} \delta u_n^{(j)})_{d} \right\} \nonumber\\
&= \exp\left\{ \delta^2 S^{(n)} (0, \delta \tilde{u}_1, \dots, \delta \tilde{u}_{n-1}, 0) \right\} \int_{\RR^{d-d'}} \dd^{d-d'} \beta \times \nonumber\\
&\quad \times \exp \left\{- \frac{1}{2} \sum_{i,j = 1}^{d-d'} \beta_i \beta_j \left( \delta u^{(i)}(0), \delta p^{(j)}(0) +(\nabla \nabla O(u_{I}(0)), {\cal F}_I)_{d'} \delta u^{(j)}(0) \right)_d \right\}\nonumber\\
&= (2 \pi)^{(d-d')/2} \left( \det B \right)^{-1/2}  \exp\left\{- \delta^2 S^{(n)} (0, \delta \tilde{u}_1, \dots, \delta \tilde{u}_{n-1}, 0) \right\}\,,
\end{align}
where we abbreviate the $(d-d')\times(d-d')$-dimensional matrix $B$ with
\begin{equation}
\label{eq:det-b}
B_{ij} := \left(\delta u^{(i)}(0), \delta p^{(j)}(0) + (\nabla \nabla O(u_{I}(0)), {\cal F}_I)_{d'} \delta u^{(j)}(0) \right)_{d,ij}\,.
\end{equation}
Summing up, at the cost of having to solve $(d-d')$ linear boundary
value problems of the form~(\ref{eq:bvp}) for each of the basis
vectors of an arbitrary orthonormal basis of $\ker \nabla O(u_I(t=0))
\subset \RR^d$ and consequently evaluating the
$(d-d')\times(d-d')$-dimensional determinant $\det B$, we are left
only with Dirichlet 0 boundary conditions $\delta u_0 = 0$ and $\delta
u_n = 0$ in the path
integral~(\ref{eq:pdf-discrete-u-small-noise}). The expression for the
PDF becomes
\begin{align}
\label{eq:intermediate-result-bvp}
&\rho_O(a) = (2 \pi)^{(d-d')/2} \eps^{-d'/2} \left[ \det B \det \left( \nabla O(u_{I,n}) \nabla O(u_{I,n})^\top \right) \right]^{-1/2} \times \nonumber\\
&\times\exp \left\{ \alpha \int_{-T}^0 \trace \left[ \nabla N(u_I(t)) \right] \dd t \right\} \exp \left\{- \eps^{-1} S_I(a) \right\} \lim_{n \to \infty} (2 \pi \Delta t)^{-nd/2}\times \nonumber\\
&\times  (\det \chi)^{-n/2} \int_{\RR^d} \left(\prod_{i=1}^{n-1} \dd^d(\delta u_i) \right)  \exp\left\{- \delta^2 S^{(n)} (0, \delta u_1, \dots, \delta u_{n-1}, 0) \right\}\,,
\end{align}
where the discrete second variation of the action is given
by~(\ref{eq:second-variation-discrete}) and evaluated with 0 boundary
conditions. Now, we turn to the computation of this remaining integral
in the continuum limit. A different approach to avoid having to solve
boundary value problems will be discussed afterwards, since, e.g.\ for
the practically relevant case of a large number of spatial dimensions
$d$ and a one dimensional observable, it is clearly undesirable to
solve $d-1$ BVPs at each $a$ where the PDF should be evaluated.

\subsection{Calculating the fluctuation determinant with Dirichlet 0 boundary conditions}
\label{subsec:fluctuation-determinant-0bc}
The computation of Gaussian path integrals with Dirichlet boundary
conditions such as the one in~(\ref{eq:intermediate-result-bvp}) which
we follow here is standard and has been discussed in many textbooks
and articles. Historically, it goes back to the works of Cameron and
Martin \cite{cameron-martin:1944} and Montroll \cite{montroll:1952}
and has been popularized in the context of one-dimensional quantum
mechanics by Gel'fand and Yaglom \cite{gelfand-yaglom:1960}. The
general $d$-dimensional case has been treated by Papadopoulos
\cite{papadopoulos:1975} and later multiple times in specific
applications, e.g.\ by Braun and Garg \cite{braun-garg:2007} or
Daumont, Dombre and Gilson \cite{daumont-dombre-gilson:2000}. Here,
however, we explicitly keep a general $\alpha$ instead of the
mid-point or Stratonovich choice $\alpha = 1/2$ in order to
demonstrate the discretization dependence of the result of the limit
in the second line of ~(\ref{eq:intermediate-result-bvp}), which is
only cured by the Jacobian that also depends on the
discretization. The discretization dependence of the determinant of
finite difference operators in the continuum limit has also been
noted, but not analyzed in detail, by Forman
\cite{forman:1992}. Furthermore, Wissel also derived
discretization-dependent Gel'fand-Yaglom formulas for the special case
of a one-dimensional Ornstein-Uhlenbeck process \cite{wissel:1979}.
The $\alpha = 0$ case of our intermediate result~(\ref{eq:gy-y}) 
has also been derived in~\cite{lehmann-reimann-haenggi:2003}.\\

The integral in~(\ref{eq:intermediate-result-bvp}) which we want to
compute in this section is
\begin{align}
I^{(n),\alpha} &= (2 \pi \Delta t)^{-nd/2} (\det \chi)^{-n/2} \times \nonumber\\
&\quad \times \int_{\RR^d} \left(\prod_{i=1}^{n-1} \dd^d(\delta u_i) \right)  \exp\left\{- \delta^2 S^{(n)} (0, \delta u_1, \dots, \delta u_{n-1}, 0) \right\}\,,
\end{align}
in the continuum limit $n \to \infty$. Substituting $\delta u_i = \sqrt{2 \Delta t} \delta \tilde{u}_i$ for $i = 1, \dots, n-1$, this integral can be expressed as
\begin{align}
\label{eq:int-dirichlet}
I^{(n),\alpha} &= (2 \pi \Delta t)^{-d/2} (\det \chi)^{-n/2} \pi^{-(n-1)d/2} \times \nonumber\\
&\quad \times \int_{\RR^d} \left(\prod_{i=1}^{n-1} \dd^d(\delta u_i) \right)  \exp\left\{- (\delta u, H^{(n-1), \alpha} \delta u)_{(n-1)d} \right\}\,,
\end{align}
where the $(n-1) d \times (n-1) d$ block tridiagonal matrix $H^{(n-1),
  \alpha}$ that can be obtained
from~(\ref{eq:second-variation-discrete}) is given by
\begin{align}
\label{eq:h-matrix-diagonal}
H^{(n-1),\alpha}_{ii} &= 2 \chi^{-1} + \Delta t (2  \alpha - 1) \left[\nabla N_i^\top \chi^{-1} + \chi^{-1} \nabla N_i \right] \nonumber\\
&\quad + \Delta t^2 \left[(\alpha^2 + (1 - \alpha)^2) \nabla N_i^\top \chi^{-1} \nabla N_i + (\nabla \nabla N_i, p_i)_d \right] \nonumber\\
&=: 2 \chi^{-1} + \Delta t R_i + \Delta t^2 S_i\,,
\end{align}
for $i = 1, \dots, n-1$ on the block diagonal (where $\nabla N_i := \nabla N(u_{I,i})$ and so on) and
\begin{align}
\label{eq:h-matrix-offdiagonal}
H^{(n-1),\alpha}_{i,i+1} &= - \chi^{-1} + \Delta t \left[ (1-\alpha) \nabla N^\top_i \chi^{-1} - \alpha \chi^{-1} \nabla N_{i+1} \right] + \Delta t^2 \alpha (1-\alpha) \nabla N_i^\top \chi^{-1} \nabla N_{i+1} \nonumber\\
&=: - \chi^{-1}  + \Delta t P^\top_i + \Delta t^2 Q^\top_i\,,
\end{align}
as well as
\begin{align}
H^{(n-1),\alpha}_{i+1, i} = -\chi^{-1} + \Delta t P_i + \Delta t^2 Q_i\,,
\end{align}
for $i = 1, \dots, n-2$. In principle, the
integral~(\ref{eq:int-dirichlet}) could be evaluated numerically by
brute force methods, either by simply computing the determinant of the
matrix $H$ numerically for large enough $n$, or by Monte Carlo
simulations, as detailed in section
\ref{subsec:continuum-overview}. We will follow both strategies for
comparison purposes in our numerical examples in section
\ref{sec:examples}. However, it is immediately clear that a direct
numerical calculation of the determinant of $H^{(n-1),\alpha}$ soon
becomes prohibitively expensive, in particular for a large number of
dimensions $d$ which one encounters when applying the formalism that
we developed here to spatially discretized partial differential
equations (even though the sparsity and structure of the block tridiagonal
matrix $H^{(n-1),\alpha}$ could in principle be exploited here).
On the other hand, a Monte Carlo approach is typically slow
and provides no analytical insights into the form and contribution of
the fluctuations around the instanton. As such, an efficient way to
evaluate~(\ref{eq:int-dirichlet}) is needed, and this is conveniently
provided by formulas of Gel'fand-Yaglom type.\\

Here, we follow the notation and derivation strategy of Ossipov
\cite{ossipov:2018} in order to derive a Gel'fand-Yaglom like,
$\alpha$-dependent equation for $I^{(n), \alpha}$ in the limit $n \to
\infty$. The basic idea can be explained quickly: We integrate out all
$\delta u_i$ step by step in chronological order. By demanding that
the result should be a Gaussian function at each step, we can then
obtain recursion relations for the parameters of these Gaussians,
which turn into a differential equation in the limit $n \to
\infty$. Hence, define
\begin{align}
\Phi_1(x) = \exp\left\{- \left(x, \left[ \chi^{-1} + \Delta t R_1 + (\Delta t)^2 S_1 \right] x \right)_d\right\}\,,
\end{align}
as well as
\begin{align}
\label{eq:gaussian-recursive}
\Phi_{k+1}(x) &= \exp \left\{ - \left( x, \left[ \Delta t R_{k+1} + \Delta t^2 S_{k+1} \right] x \right)_d\right\} \pi^{-d/2} \int_{\RR^d} \dd^d y \; \times \nonumber\\
&\quad \times \exp \left\{ -(x-y, \chi^{-1} [x-y] )_d - 2 \left( x, \left[ \Delta t P_k + \Delta t^2 Q_k \right] y \right)_d \right\} \Phi_k(y)\,,
\end{align}
for $k = 1, \dots, n-1$. Then, we can express $I^{(n),\alpha}$ as
\begin{equation}
I^{(n),\alpha} = (2 \pi \Delta t)^{-d/2} (\det \chi)^{-n/2} \Phi_n(0)\,.
\end{equation}
Now, we insert the general Gaussian ansatz
\begin{equation}
\label{eq:gaussian-recursive-ansatz}
\Phi_k(x) = c_k \exp\left\{-(x, A_k x)_d - (b_k,x) \right\}\,,
\end{equation}
with parameters $c_k > 0$, $A_k \in \RR^{d \times d}$ symmetric and positive definite, and $b_k \in \RR^d$. Clearly, we have
\begin{equation}
A_1 = \chi^{-1} + \Delta t R_1 + \Delta t^2 S_1, \quad b_1 = 0,\quad c_1 = 1\,,
\end{equation}
as initial values for these parameters. Plugging in the
ansatz~(\ref{eq:gaussian-recursive-ansatz})
into~(\ref{eq:gaussian-recursive}) yields the recursion relations
\begin{align}
\label{eq:recursion-a}
A_{k+1} =& \chi^{-1} + \Delta t R_{k+1} + \Delta t^2 S_{k+1} \nonumber\\
&\quad - \left(\chi^{-1} - \Delta t P_{k} - \Delta t^2 Q_k \right) \left(\chi^{-1} + A_k \right)^{-1} \left(\chi^{-1} - \Delta t P_k^\top - \Delta t^2 Q_k^\top \right)\,,
\end{align}
(which is also true for $k=0$ if we define $A_0 = \infty$) as well as
\begin{equation}
b_{k+1} = \left(\chi^{-1} - \Delta t P_k - \Delta t^2 Q_k \right) \left(\chi^{-1} + A_k\right)^{-1} b_k\,,
\end{equation}
and
\begin{equation}
c_{k+1} = c_k \left[\det\left( \chi^{-1} + A_k \right) \right]^{-1/2} \exp\left\{\frac{1}{4} \left(b_k, \left(\chi^{-1} + A_k \right)^{-1} b_k \right)_d \right\}\,.
\end{equation}
All of these relation directly follow from applying to~(\ref{eq:gaussian-recursive}) the general identity
\begin{equation}
\label{eq:general-gaussian}
\int_{\RR^d} \dd^dx\; \exp \left\{-(x,Ax)_d + (b,x)_d \right\} = \left[ \det \left(\frac{A}{\pi} \right) \right]^{-1/2} \exp \left\{\frac{1}{4} \left(b, A^{-1} b \right)_d \right\}
\end{equation}
for a Gaussian integral with source term, which we explicitly state
here for later convenience. Since $b_1 = 0$, we immediately obtain
$b_k = 0$ for all $k = 1, \dots, n$, such that
\begin{align}
I^{(n),\alpha} &= (2 \pi \Delta t)^{-d/2} (\det \chi)^{-n/2} \Phi_n(0) =  (2 \pi \Delta t)^{-d/2} (\det \chi)^{-n/2} c_n \nonumber\\
&= (2 \pi)^{-d/2} \left[ \Delta t^d (\det \chi)^n \prod_{k=1}^{n-1} \det \left( \chi^{-1} + A_k \right) \right]^{-1/2}\,.
\end{align}
Now, we define
\begin{equation}
\chi^{-1} + A_k =: \chi^{-1} Y_{k+1} Y_k^{-1}, \quad k = 1, \dots, n-1\,.
\end{equation}
With $A_0 = \infty$, we set $Y_0 = 0 \in \RR^{d \times d}$, and we are free to choose $Y_1$. Taking
\begin{equation}
  Y_1 = \Delta t \chi\,,
\end{equation}
the integral $I^{(n), \alpha}$ simply becomes
\begin{equation}
I^{(n),\alpha} = (2 \pi)^{-d/2} \left[ \det Y_n \right]^{-1/2}\,,
\end{equation}
with this ansatz. The quantities $(Y_k)$ do in fact possess a
well-defined continuum limit, since we absorbed all remaining
divergent constants in their definition. It is obvious that the
initial values for the continuum limit $Y(t)$ will be
\begin{equation}
  Y(-T) = 0, \quad \dot{Y}(-T) = \chi\,.
\end{equation}
As for the recursion relation in terms of $Y$,~(\ref{eq:recursion-a}) yields
\begin{align}
A_{k+1} &= \chi^{-1} Y_{k+2} Y_{k+1}^{-1} - \chi^{-1} \nonumber\\
&= \chi^{-1} + \Delta t R_{k+1} + (\Delta t)^2 S_{k+1} \nonumber\\
&\quad - \left( \chi^{-1} - \Delta t P_k - \Delta t^2 Q_k \right) Y_k Y_{k+1}^{-1} \chi  \left( \chi^{-1} - \Delta t P_k^\top - \Delta t^2 Q_k^\top \right)\,,
\end{align}
or, sorting by powers of $\Delta t$ and ignoring terms that will vanish for $\Delta t \to 0$:
\begin{align}
\label{eq:recursion-y}
&\chi^{-1} \frac{Y_{k+2} - 2 Y_{k+1} + Y_k}{\Delta t^2} - \frac{1}{\Delta t} \left[ R_{k+1} Y_{k+1} + P_k Y_k + \chi^{-1} Y_k Y_{k+1}^{-1} \chi P_k^\top Y_{k+1} \right] \nonumber \\
&- \left[ S_{k+1} Y_{k+1} + Q_k Y_k + \chi^{-1} Y_k Y_{k+1}^{-1} \chi Q_k^\top Y_{k+1} - P_k Y_k Y_{k+1}^{-1} \chi P_{k}^\top Y_{k+1} \right] = 0\,.
\end{align}
The first term clearly converges to $\chi^{-1} \ddot{Y}$ in the
continuum limit, but the other two terms require a more careful
treatment. The second term in the first line of~(\ref{eq:recursion-y})
is given by
\begin{align}
&- \frac{1}{\Delta t} \left[ (2 \alpha - 1) \left( \nabla N_{k+1}^\top \chi^{-1} + \chi^{-1} \nabla N_{k+1} \right) Y_{k+1} + \left( (1-\alpha) \chi^{-1} \nabla N_k - \alpha \nabla N_{k+1}^\top \chi^{-1} \right) Y_k \right.\nonumber\\
&\left.  + \chi^{-1} Y_k Y_{k+1}^{-1} \chi \left( (1-\alpha) \nabla N_k^\top \chi^{-1} - \alpha \chi^{-1} \nabla N_{k+1} \right) Y_{k+1} \right]\,.
\end{align}
Now, we expand
\begin{align}
Y_{k+1} = Y_k + \Delta t \frac{Y_{k+1} - Y_k}{\Delta t} = Y_k + \Delta t \dot{Y}_k\,,
\end{align}
and use
\begin{align}
\frac{\dd}{\dd t} Y^{-1} = - Y^{-1} \dot{Y} Y^{-1}\,,
\end{align}
such that
\begin{align}
Y_{k+1}^{-1} = Y_k^{-1} - \Delta t Y_k^{-1} \dot{Y}_k Y_k^{-1}\,.
\end{align}
Inserting these expansions yields the following continuum limit for this term:
\begin{align}
&(1-\alpha) \chi^{-1} \frac{\dd}{\dd t} \left( \nabla N Y \right) + (1-\alpha) \frac{\dd}{\dd t} \left( \nabla N^\top \right) \chi^{-1} Y - \alpha \nabla N^\top \chi^{-1}\dot{Y} \nonumber\\
&+ (1-\alpha) \chi^{-1} \dot{Y} Y^{-1} \chi \nabla N^\top \chi^{-1} Y - \alpha \chi^{-1} \dot{Y} Y^{-1} \nabla N Y\,.
\end{align}
The remaining terms of~(\ref{eq:recursion-y}) are of order 1 in $\Delta t$, their limit is
\begin{align}
&(\alpha^2 - 1) \nabla N^\top \chi^{-1} \nabla N Y - \left( \nabla \nabla N,p_I \right)_d Y + (1-\alpha)^2 \chi^{-1} \nabla N \chi \nabla N^\top \chi^{-1} Y \nonumber\\
&- \alpha (1-\alpha) \left[ \chi^{-1} \nabla N^2 + \left( \nabla N^\top \right)^2 \chi^{-1}\right] Y\,.
\end{align}
Summing up, we arrive at the final result
\begin{align}
\lim_{n \to \infty} I^{(n),\alpha} = (2 \pi)^{d/2} \left[\det Y(0)\right]^{-1/2}\,,
\end{align}
for the path integral, where $Y \in \RR^{d \times d}$ solves the (in
general, for $d > 1$) nonlinear second order matrix differential
equation
\begin{align}
\label{eq:gy-y}
&\chi^{-1} \ddot{Y} + (1-\alpha) \chi^{-1} \frac{\dd}{\dd t} \left( \nabla N Y \right) + (1-\alpha) \frac{\dd}{\dd t} \left( \nabla N^\top \right) \chi^{-1} Y - \alpha \nabla N^\top \chi^{-1} \dot{Y} \nonumber\\
&+ (1-\alpha) \chi^{-1} \dot{Y} Y^{-1} \chi \nabla N^\top \chi^{-1} Y - \alpha \chi^{-1} \dot{Y} Y^{-1} \nabla N Y \nonumber \\
&+ (\alpha^2 - 1) \nabla N^\top \chi^{-1} \nabla N Y - \left( \nabla \nabla N,p_I \right)_d Y + (1-\alpha)^2 \chi^{-1} \nabla N \chi \nabla N^\top \chi^{-1} Y \nonumber\\
&- \alpha (1-\alpha) \left[ \chi^{-1} \nabla N^2 + \left( \nabla N^\top \right)^2 \chi^{-1}\right] Y = 0\,,
\end{align}
with initial conditions $Y(-T) = 0, \dot{Y}(-T) = \chi$. This unwieldy
equation does in fact depend on $\alpha$, and so does the value of
$\lim_{n \to \infty} I^{(n),\alpha}$, but we are free to choose any
$\alpha$ from now on in order to bring this equation into a simpler
form. Obviously, the choice $\alpha = 1$, which corresponds to a fully
implicit discretization of the SDE~(\ref{eq:discretized-sde}), is
advantageous as most terms of~(\ref{eq:gy-y}) will vanish in this
case. This leads to
\begin{align}
\label{eq:gy-y-alpha-1}
\chi^{-1} \ddot{Y} - \nabla N^\top \chi^{-1} \dot{Y}- \chi^{-1} \dot{Y} Y^{-1} \nabla N Y - \left( \nabla \nabla N,p_I \right)_d Y = 0\,,
\end{align}
which is still nonlinear, but can be transformed into a symmetric matrix Riccati
differential equation for which there exist well-known solution methods
(see \cite{benner-mena:2004} for an overview). Indeed,
setting $Q = Y \dot{Y}^{-1} \chi$, we obtain
\begin{align}
\label{eq:riccati-alpha-1}
\dot{Q} = \chi - Q \nabla N^\top - \nabla N Q - Q (\nabla \nabla N,p_I)_d Q, \quad Q(-T) = 0 \in \RR^{d \times d}\,.
\end{align}
Depending on the system at hand, it can be numerically or
theoretically advantageous to linearize~(\ref{eq:riccati-alpha-1}) by
a Radon transform \cite{radon:1928}: Defining $Q = \delta U \delta
P^{-1}$ with $\delta U,\; \delta P \in \RR^{d \times d}$, we have
$\delta U(-T) = 0$ and are free to choose $\delta P(-T) = 1$. Then, by
demanding that these matrices satisfy a linear matrix differential
equation
\begin{align}
\frac{\dd}{\dd t} \left( \begin{array}{c}
\delta U\\
\delta P
\end{array} \right) = \left( \begin{array}{cc}
M_{11} & M_{12}\\
M_{21} & M_{22}
\end{array} \right) \left( \begin{array}{c}
\delta U\\
\delta P
\end{array} \right)\,,
\end{align}
and inserting the ansatz into~(\ref{eq:riccati-alpha-1}), we obtain
\begin{align}
\label{eq:radon}
\frac{\dd}{\dd t}  \left( \begin{array}{c}
\delta U\\
\delta P
\end{array} \right) = \left( \begin{array}{cc}
- \nabla N & \chi\\
(\nabla \nabla N,p_I)_d & \nabla N^\top
\end{array} \right) \left( \begin{array}{c}
\delta U\\
\delta P
\end{array} \right), \; \delta U(-T) = 0, \; \delta P(-T) = 1\,.
\end{align}
Remarkably, by these transformations we obtain a classical, linear
Gel'fand-Yaglom formula that is equivalent to $H \delta u = 0$ where
$H$ is given by~(\ref{eq:second-variation-continuum}), which occurs as
a matrix-valued linear first order initial value problem (IVP) in this
case and was obtained for the choice of $\alpha = 1$, and not $\alpha
= 1/2$. However, we note that linearizing the Riccati equation by this
substitution may not be advisable numerically, since the $\delta P$
equation in~(\ref{eq:radon}) is integrated forward in time in this
case, but the term $\nabla N^\top \delta P$ on the right-hand side
of~(\ref{eq:radon}) has a different sign than the drift term in the
original SDE~(\ref{eq:SDE}). Hence, if the original system is
dissipative, the amplitude of $\delta P$, and consequently, since it
occurs as a forcing term in the respective equation, also the
amplitude of $\delta U$ will grow exponentially in time. The nonlinear
Riccati equation~(\ref{eq:riccati-alpha-1}) does not possess this
property, but its nonlinearity is undesirable in the sense that for
very large dimensions $d$, as would be encountered in the spatial
discretization of multi-dimensional PDEs, the solution of the linear
equation~(\ref{eq:radon}) could be \textit{parallelized} trivially
over the column vectors of $\delta U$ and $\delta P$.\\

In order to be able to express our final result for the PDF $\rho_O$
in the second order expansion in terms of the solutions of the
BVPs~(\ref{eq:bvp}) and the Riccati IVP~(\ref{eq:riccati-alpha-1}), we
still have to express
\begin{equation}
\label{eq:gy-result}
\lim_{n \to \infty} I^{(n), 1} = (2 \pi)^{-d/2} \left[ \det Y(0) \right]^{-1/2} = (2 \pi)^{-d/2} \left[ \det Q(0) (\det \chi)^{-1} \det \dot{Y}(0) \right]^{-1/2}
\end{equation}
fully in terms of $Q$. In order to do this, we calculate
\begin{align}
&(\det \chi)^{-1} \det \dot{Y}(0) = \frac{\det \dot{Y}(0)}{\det \dot{Y}(-T)} = \exp\left\{\trace \log \dot{Y}(0) - \trace \log \dot{Y}(-T)\right\}\nonumber\\
&= \exp \left\{ \int_{-T}^0 \dd t \frac{\dd}{\dd t} \; \left(\trace \log \dot{Y} \right)\right\} = \exp \left\{ \int_{-T}^0 \dd t \; \trace \left[ \ddot{Y} \dot{Y}^{-1} \right] \right\} \nonumber\\
&\overset{(\ref{eq:gy-y})}{=} \exp \left\{ \int_{-T}^0 \dd t \; \trace \left[ \chi \left(  \nabla N^\top \chi^{-1} \dot{Y} + \chi^{-1} \dot{Y}Y^{-1} \nabla N Y + (\nabla \nabla N,p_I)_d Y\right) \dot{Y}^{-1} \right] \right\} \nonumber\\
&=\exp \left\{\int_{-T}^0 \dd t\; \trace \left[2 \nabla N + (\nabla \nabla N, p_I)_d Q \right]\right\}\,,
\label{eq:exp-tr-log-y-to-w}
\end{align}
where we repeatedly used the cyclicity property of the trace in the
last line, as well as in the differentiation in the second line in
order to be able to differentiate $\log \dot Y$ as if it was a
scalar. Putting everything together we obtain the following final
expression for the PDF $\rho_O$ of a nonlinear, $d'$-dimensional
observable of the stochastic process described by the $d$-dimensional
SDE~(\ref{eq:SDE}) in the small noise and continuum limit:
\begin{align}
\label{eq:final-result-including-bvp}
\rho_O(a) &= (2 \pi \eps)^{-d'/2} \left[ \det B \det \left( \nabla O(u_{I}(0)) \nabla O(u_{I}(0))^\top \right) \right]^{-1/2} \left(\det Q(0)\right)^{-1/2} \times \nonumber\\
& \times \exp \left\{- \frac{1}{2} \int_{-T}^0 \dd t \; \trace \left[(\nabla \nabla N(u_I(t)), p_I(t))_d Q(t) \right]  \right\} \exp\left\{-\eps^{-1} S_I(a) \right\}\,.
\end{align}
The expression which is shown here was derived for $\alpha = 1$ since
this choice clearly yields the simplest result based on our previous
discussion. 
To summarize what has been discussed so far, the method which we just
introduced consists of three major steps in order to evaluate the
complete second order approximation to the PDF $\rho_O$ at each $a \in
\RR^{d'}$:
\begin{enumerate}
\item Calculate the instanton trajectory $(u_I, p_I)$, which is the
  solution of the minimization
  problem~(\ref{eq:minimization-problem}). The observable value $a$
  implicitly enters as a boundary condition, leading to an
  $a$-dependent action $S_I(a)$ at the instanton that determines the
  ${\cal O}(e^{\eps^{-1}})$ contribution to the PDF. The instanton
  then enters as a background field into the differential equations
  that need to be solved for the prefactor, and thus introduces
  $a$-dependence into the prefactor.\label{item:numerics-inst}
\item Solve $d-d'$ boundary value problems~(\ref{eq:bvp}), and
  evaluate the final time contribution $\det B$ of their
  solutions.\label{item:numerics-BVP}
\item Solve a matrix Riccati equation~(\ref{eq:riccati-alpha-1}) as an
  initial value problem for $Q$ and evaluate the corresponding
  integral in~(\ref{eq:final-result-including-bvp}) along the
  trajectory as well as the determinant of
  $Q(t=0)$.\label{item:numerics-Riccati}
\end{enumerate}
We want to stress at this point that even though a consistent
discretization was crucial in the derivation
of~(\ref{eq:final-result-including-bvp}), all points
\ref{item:numerics-inst} to \ref{item:numerics-Riccati} from the list
given above can numerically be solved using any discretization or
integration scheme that one wants to apply. In the next section, we
turn to a simpler alternative to~(\ref{eq:final-result-including-bvp})
that circumvents the possibly large number of boundary value problems
in the method outlined so far, and can be derived quickly from our
previous discussion.

\subsection{Alternative approach without homogenization}
\label{subsec:prefac-efficient}

While the reduction of the boundary conditions to Dirichlet $0$ is
desirable from a theoretical point of view in order to be able to
connect our result to other studies that evaluate functional
determinants for differential operators with such boundary conditions,
the necessity to solve a number of boundary value problems which
scales linearly with the system dimension $d$ (if a one-dimensional
observable, $d'=1$, is considered) is clearly undesirable from a
practical and in particular numerical point of view. Hence, we will
derive an alternative, much simpler approach to evaluate the prefactor
in this section that does not require the solution of boundary value
problems. In fact, the solution of the Riccati
equation~(\ref{eq:riccati-alpha-1}) already contains all necessary
information to evaluate the prefactor. In this section, we will
directly work with the $\alpha = 1$ discretization which was shown to
be the optimal choice in the previous section.\\

Our starting point is the Gaussian
integral~(\ref{eq:pdf-discrete-u-small-noise}) for $\alpha = 1$
\begin{align}
 \rho_O(a) &= \eps^{-d'/2} \lim_{n \to \infty} \left(2 \pi \Delta t \right)^{-nd/2} \left( \det \chi\right)^{-n/2} \exp \left\{\int_{-T}^0 \trace \left[ \nabla N(u_I(t)) \right] \dd t \right\} \exp \left\{- \eps^{-1} S_I(a) \right\} \times \nonumber\\
 & \times \int_{\RR^d} \left( \prod_{j=1}^{n} \dd^d (\delta u_j) \right) \delta(\nabla O(u_{I,n})\delta u_n) \times \nonumber\\
 &\times \exp\left\{- \delta^2 S^{(n)}[\delta u] - \frac{1}{2}(\delta u_n, (\nabla \nabla O(u_{I,n}), {\cal F}_I)_{d'} \delta u_n)_{d}\right\}\,,
\end{align}
with
\begin{align}
\delta^2 S^{(n)}[\delta u] &= \frac{\Delta t}{2} \bigg( \sum_{i=0}^{n-1} \bigg[ \left(\frac{ \delta u_{i+1} - \delta u_i}{\Delta t} + \nabla N(u_{I,i+1}) \delta u_{i+1}, \chi^{-1} \left[\dots  \right] \right)_d \nonumber\\
& + \left(\delta u_i, \left(\nabla \nabla N(u_{I,i}), p_{I,i} \right)_d \delta u_{i} \right)_d \bigg] + \left(\delta u_n, \left(\nabla \nabla N(u_{I,n}), p_{I,n} \right)_d \delta u_{n} \right)_d \bigg)\,.
\end{align}
Again substituting $\delta u_i = \sqrt{2 \Delta t} \delta \tilde{u_i}$, the PDF becomes
\begin{align}
\rho_O(a) &= \eps^{-d'/2} \exp \left\{\int_{-T}^0 \trace \left[ \nabla N(u_I(t)) \right] \dd t \right\} \exp \left\{- \eps^{-1} S_I(a) \right\} \lim_{n \to \infty} \left( \det \chi\right)^{-n/2} \times \nonumber\\
 & \times \pi^{-nd/2} \int_{\RR^d} \left( \prod_{j=1}^{n} \dd^d (\delta u_j) \right) \delta( \sqrt{2 \Delta t} \nabla O(u_{I,n}) \delta u_n) \times\nonumber\\
 &\times \exp\left\{- (\delta u, H ^{(n), 1}\delta u)_{nd} - \frac{1}{2}(\delta u_n, (\nabla \nabla O(u_{I,n}), {\cal F}_I)_{d'} \delta u_n)_{d}\right\}\,,
\end{align}
with the symmetric $nd \times nd$ block tridiagonal matrix $H ^{(n),
  1}$ with diagonal entries
\begin{equation}
  H ^{(n), 1}_{ii} = 2 \chi^{-1} + \Delta t \left[ \nabla N_i^\top \chi^{-1} + \chi^{-1} \nabla N_i \right] + \Delta t^2 \left[\nabla N_i^\top \chi^{-1} \nabla N_i + (\nabla \nabla N_i, p_i)_d \right]\,,
\end{equation}
for $i=1, \dots, n-1$, but
\begin{equation}
  H ^{(n), 1}_{nn} = \chi^{-1} + \Delta t \left[ \nabla N_n^\top \chi^{-1} + \chi^{-1} \nabla N_n \right] + \Delta t^2 \left[\nabla N_n^\top \chi^{-1} \nabla N_n + (\nabla \nabla N_n, p_n)_d \right]\,.
\end{equation}
The off-diagonal nonzero blocks are
\begin{equation}
  H ^{(n), 1}_{i, i+1} = -\chi^{-1} - \Delta t \chi^{-1} \nabla N_{i+1} = H_{i+1,i}^\top\,,
\end{equation}
for $i = 1, \dots, n-1$. Now, the key observation is that the
additional $\delta u_n$ integral that occurs here does not interfere
with the way in which we derived the recursion relation for the
sequence of Gaussian integral in the previous section. Instead, using
the same nomenclature as in section
\ref{subsec:fluctuation-determinant-0bc}, the PDF can be written as
\begin{align}
\rho_O(a) &= \eps^{-d'/2} \exp \left\{\int_{-T}^0 \trace \left[ \nabla N(u_I(t)) \right] \dd t \right\} \exp \left\{- \eps^{-1} S_I(a) \right\} \times \nonumber\\
 & \times \lim_{n \to \infty} \left( \det \chi\right)^{-n/2} \pi^{-d/2} \int_{\RR^d} \dd^d (\delta u_n) \, \delta( \sqrt{2 \Delta t} \nabla O(u_{I,n}) \delta u_n) \times \nonumber\\
 &\times \exp\left\{- \Delta t(\delta u_n, (\nabla \nabla O(u_{I,n}), {\cal F}_I)_{d'} \delta u_n)_{d}\right\}  \Phi_n(\delta u_n)\,,
\end{align}
with the function
\begin{equation}
  \Phi_n(\delta u_n) = c_n \exp \left\{-(\delta u_n, A_n \delta u_n) \right\}\,,
\end{equation}
resulting from recursive Gaussian integration as discussed previously.
Again using~(\ref{eq:delta-fourier}) for the $d'$-dimensional
$\delta$-function, this can be rewritten as
\begin{align}
\label{eq:pdf-efficient-prelim}
\rho_O(a) &= (2 \pi \eps)^{-d'/2} \exp \left\{\int_{-T}^0 \trace \left[ \nabla N(u_I(t)) \right] \dd t \right\} \exp \left\{- \eps^{-1} S_I(a) \right\} \times \nonumber\\
 & \times \lim_{n \to \infty} \left( \det \chi\right)^{-n/2} c_n (2 \pi)^{-d'/2} \int_{\RR^{d'}} \dd^{d'} k \; \pi^{-d/2} \int_{\RR^d} \dd^d (\delta u_n) \times  \nonumber\\
 &\times \exp\left\{ -(\delta u_n, \left[ A_n + \Delta t (\nabla \nabla O(u_{I,n}),{\cal F}_I)_{d'} \right] \delta u_n) + \sqrt{2 \Delta t} i (\nabla O(u_{I,n})^\top k, \delta u_n)_d \right\}\,.
\end{align}
The general formula~(\ref{eq:general-gaussian}) for Gaussian integrals with source term shows that the last $\delta u_n$-integral in ~(\ref{eq:pdf-efficient-prelim}) evaluates to
\begin{align}
  &\pi^{-d/2} \int_{\RR^d} \dd^d (\delta u_n) \, \exp\left\{ -(\delta u_n, \left[ A_n + \Delta t (\nabla \nabla O(u_{I,n}),{\cal F}_I)_{d'} \right] \delta u_n) \right\} \times\nonumber\\
   &\times \exp \left\{ \sqrt{2 \Delta t} i (\nabla O(u_{I,n})^\top k, \delta u_n)_d \right\} \nonumber\\
  &= \left[ \det  \left( A_n + \Delta t (\nabla \nabla O(u_{I,n}),{\cal F}_I)_{d'}\right) \right]^{-1/2} \times \nonumber\\
  & \times \exp \left\{ - \frac{1}{2} \left(\nabla O(u_{I,n})^\top k, \left(\frac{A_n}{\Delta t} + (\nabla \nabla O(u_{I,n}),{\cal F}_I)_{d'}\right)^{-1} \nabla O(u_{I,n})^\top k\right)_d  \right\}\,,
\end{align}
so the $k$-integral in ~(\ref{eq:pdf-efficient-prelim}) can also easily be evaluated and leads to the final result
\begin{align}
\label{eq:pdf-efficient-prelim2}
\rho_O(a) &= (2 \pi \eps)^{-d'/2} \exp \left\{\int_{-T}^0 \trace \left[ \nabla N(u_I(t)) \right] \dd t \right\} \exp \left\{- \eps^{-1} S_I(a) \right\} \times \nonumber\\
 & \times \lim_{n \to \infty} \Bigg[ \left( \det \chi\right)^n c_n^{-2}  \det  \left( A_n + \Delta t (\nabla \nabla O(u_{I,n}),{\cal F}_I)_{d'}\right)  \times \nonumber \\
 & \times \det \left(\nabla O(u_{I,n}) \left(\frac{A_n}{\Delta t} + (\nabla \nabla O(u_{I,n}),{\cal F}_I)_{d'}\right)^{-1} \nabla O(u_{I,n})^\top \right)\Bigg]^{-1/2}\,.
\end{align}
By the definitions from section~\ref{subsec:fluctuation-determinant-0bc}, we have
\begin{align}
 &\left( \det \chi\right)^n c_n^{-2}  \det  \left( A_n + \Delta t (\nabla \nabla O(u_{I,n}),{\cal F}_I)_{d'}\right) \nonumber\\
 &= \Delta t^{-d}\det Y_n  \det  \left( A_n + \Delta t (\nabla \nabla O(u_{I,n}),{\cal F}_I)_{d'}\right)\nonumber\\
 &= \det Y_n \det \left( \frac{A_n}{\Delta t} + (\nabla \nabla O(u_{I,n}),{\cal F}_I)_{d'}\right),
\end{align}
and the continuum limit of $A_n / \Delta t$ is found from
\begin{equation}
 \chi^{-1} + A_n = \chi^{-1} Y_{n+1} Y_n^{-1} = \chi^{-1} \left( Y_n + \Delta t \dot{Y}_n + {\cal O}\left(\Delta t^2 \right) \right) Y_n^{-1}\,,
\end{equation}
such that
\begin{equation}
 \frac{A_n}{\Delta t} \xrightarrow{n \to \infty} \chi^{-1} \dot{Y}(0) Y(0)^{-1} = Q(0)^{-1}\,,
\end{equation}
and, using the same steps as in section
\ref{subsec:fluctuation-determinant-0bc} to express $\dot{Y}(0)$ in
terms of $Q$,
\begin{equation}
  \det Y_n \xrightarrow{n \to \infty} \det Q(0) \exp \left\{ \int_{-T}^0 \dd t \; \trace \left[2 \nabla N + (\nabla \nabla N, p_I)_d Q \right] \right\}\,.
\end{equation}
Plugging these limits into~(\ref{eq:pdf-efficient-prelim2}) and defining
\begin{align}
U = 1 + \left(\nabla \nabla O(u_I(0)), {\cal F}_I \right)_{d'} Q(0)\,,
\end{align}
yields the final, and central result of this paper:
\begin{align}
  \label{eq:final-result-wo-bvp}
  \rho_O(a) &= (2 \pi \eps)^{-d'/2} \exp \left\{- \frac{1}{2} \int_{-T}^0 \dd t \; \trace \left[ (\nabla \nabla N(u_I(t)), p_I(t))_d Q(t) \right] \right\} \times \nonumber\\
  & \quad \times
  \left[\det U  \det \left(\nabla O(u_{I}(0)) Q(0) U^{-1} \nabla O(u_{I}(0))^\top \right)\right]^{-1/2} \exp \left\{- \eps^{-1} S_I \right\}\,.
\end{align}
This equation estimates the complete prefactor for the PDF of a
$d'$-dimensional observable $O$ in the small noise limit in terms of
the solution $Q$ of a single matrix Riccati equation
\begin{equation}
\label{eq:riccati-w-final}
  \dot{Q} = \chi - Q \nabla N^\top(u_I) - \nabla N(u_I) Q - Q(\nabla \nabla N(u_I), p_I)_d Q, \quad Q(-T) = 0\,,
\end{equation}
that can easily be evaluated numerically once the instanton is known,
even for large system dimensions $d$.

\begin{remark} \label{rem:feynman-kac}
  It is also possible to derive~(\ref{eq:final-result-wo-bvp}) 
  and~(\ref{eq:riccati-w-final}) based solely on
  probabilistic methods, without explicit reference to the path integral
  computations that were utilized above, by adopting the techniques 
  from~\cite{dean-miao-podgornik:2019}. Starting from~(\ref{eq:prefac-expectation}),
  we note that, for suitable functions $f,g : \RR^d \to \RR$, the prefactor 
  can be written as
  \begin{align}
  Z = \bigg< f(\delta u(0)) \exp \left\{- \int_{-T}^0 \dd t \; g(\delta u(t)) \right\} \bigg>\,,
  \end{align}
  where $\delta u$ is the Gaussian process defined by~(\ref{eq:sde-linear}).
  Expectations of this form can be computed by the Feynman-Kac formula, which,
  in its forward version, states that
  \begin{align}
  Z = \int_{\RR^d} \dd^d v\; f(v) K(v, 0; 0, -T)\,. \label{eq:fk-int-prop}
  \end{align}
  Here, the propagator 
  \begin{align}
  K(v, t; \delta u(-T) = 0, -T) := \left< \delta(\delta u(t) - v) \exp\left\{ -\int_{-T}^t \dd t' \; g\left(\delta u \left(t' \right) \right) \right\} \right>_{\delta u(-T) = 0}
  \end{align} 
  solves
  \begin{align}
  \partial_t K(v, t; 0, -T) = \left[ G_v^\dagger - g(v) \right] K(v, t; 0, -T)\,,
  \label{eq:prop-fk-eq}
  \end{align}
  with $G^\dagger_v$ denoting the adjoint of the infinitesimal generator
  \begin{align}
  G_v = -\left(\nabla N(u_I) v, \nabla_v \; \cdot \; \right)_d + \frac{1}{2} \trace \left[ \chi \nabla_v \nabla_v \; \cdot \; \right]
  \end{align}
  of the process $\delta u$, and initial condition
  \begin{align}
  K(v, -T; 0, -T) = \delta(v)\,.
  \label{eq:init-cond-prop}
  \end{align}
  For the SDE~(\ref{eq:sde-linear}) and 
  $g(v) = \frac{1}{2} \left(v, \left(\nabla \nabla N(u_I),p_I \right)_d v \right)_d$, 
  the propagator equation~(\ref{eq:prop-fk-eq}) becomes
  \begin{align}
  \partial_t K = \trace \left[\nabla N(u_I) \right] K + \left( \nabla N(u_I) v, \nabla_v K \right)_d + \frac{1}{2} \trace \left[\chi \nabla_v \nabla_v K \right] - \frac{1}{2} \left(v, \left(\nabla \nabla N(u_I),p_I \right)_d v \right)_d K\,.
  \label{eq:prop-fk-eq-here}
  \end{align}
  Inserting the Gaussian ansatz
  \begin{align}
  K(v,t;0,-T) = c \exp \left\{-\mu(t) - \frac{1}{2} \left(v, Q^{-1}(t) v \right)_d \right\} \label{eq:prop-fk-ansatz}
  \end{align}
  with $\mu:[-T,0] \to \RR$ and a symmetric matrix $Q:[-T,0] \to \RR^{d \times d}$
  into~(\ref{eq:prop-fk-eq-here}) and sorting by orders of $v$ gives
  \begin{align}
  \dot{\mu} &= \frac{1}{2} \trace \left[\chi Q^{-1} - 2 \nabla N(u_I) \right]\,, \label{eq:mu-dot}\\
  \dot{Q}	&= \chi - \nabla N(u_I) Q - Q \nabla N(u_I)^\top - Q \left(\nabla \nabla N(u_I),p_I \right)_d Q\,.
  \end{align}
  We see that this ansatz immediately recovers the differential
  Riccati equation~(\ref{eq:riccati-w-final}), and the initial
  condition~(\ref{eq:init-cond-prop}) necessitates $Q(t \to -T) \to 0$.
  Integrating~(\ref{eq:mu-dot}) and proceeding as in~(\ref{eq:exp-tr-log-y-to-w}),
  we find
  \begin{align}
  \mu(t)-\mu(-T) = \frac{1}{2} \left(\trace \left[\log Q(t) \right] -\trace \left[\log Q(-T) \right] + \int_{-T}^t \dd t' \; \trace \left[\left(\nabla \nabla N(u_I),p_I \right)_d Q \right]\right)\,,
  \end{align}
  so the propagator~(\ref{eq:prop-fk-ansatz}) becomes
  \begin{align}
  K(v,t;0,-T) = \tilde{c} \; \left[\det Q(t) \right]^{-1/2} \exp \left\{ - \frac{1}{2}\int_{-T}^t \dd t' \; \trace \left[\left(\nabla \nabla N(u_I),p_I \right)_d Q \right] - \frac{1}{2} \left(v, Q^{-1}(t) v \right)_d \right\}\,,
  \end{align}
  where all constants were absorbed into $\tilde{c}$, which, due to the initial
  condition~(\ref{eq:init-cond-prop}), turns out to be $\tilde{c} = (2 \pi)^{-d/2}$.
  With this expression for $K$, we obtain from~(\ref{eq:fk-int-prop}):
  \begin{align}
  Z &= (2 \pi)^{-d/2} \eps^{-d'/2} \left[\det Q(0) \right]^{-1/2} \exp \left\{ - \frac{1}{2}\int_{-T}^0 \dd t' \; \trace \left[\left(\nabla \nabla N(u_I),p_I \right)_d Q \right] \right\} \times \nonumber\\
  &\quad \times \int_{\RR^d} \dd^d v \; \delta(\nabla O(u_I(0)) v) \exp\left\{-\frac{1}{2} \left(v, \left[Q^{-1}(0) + \left(\nabla \nabla O(u_I(0)), {\mathcal F_I} \right)_{d'} \right] v \right)_d \right\}\,.
  \end{align}    
  The Gaussian integral in the second line can easily be evaluated
  analogously to the computations in the previous section, and this
  precisely reproduces~(\ref{eq:final-result-wo-bvp}). We also remark
  that this prefactor computation method based on the Feynman-Kac
  equation could immediately be generalized to include higher order
  fluctuations as discussed for example in~\cite{kifer:1977,bouchet-gawedzki-nardini:2016,ferre-grafke:2020}.
\end{remark}

\section{Examples}
\label{sec:examples}

In this section we show two examples of low-dimensional SDEs as a proof of
concept for the prefactor computation strategy that we developed in
the previous section, as well as preliminary results for the application
to the stochastic Burgers equation in one spatial dimension.
The detailed analysis of the prefactor computation strategy and its results
for the Burgers equation and other SPDEs will be the subject of separate, future work.

\subsection{One-dimensional gradient system}
We start with the example of a one-dimensional SDE
\begin{equation}
\label{eq:SDE-1d}
  \dot{u} + V'(u) = \eta, \quad \left<\eta(t) \eta(t') \right> = 2 \eps \delta(t-t')\,,
\end{equation}
where $V:\RR \to \RR$ is a smooth potential with a unique, stable and
non-degenerate fixed point $\bar{x} \in \RR$, such that $V'(\bar{x}) =
0$ and $V''(\bar{x}) > 0$. We consider the SDE~(\ref{eq:SDE-1d}) on
the time interval $[-T,0]$ with deterministic initial condition $u(-T)
= \bar{x}$, such that the process starts at the fixed point of the
dynamics. We want to evaluate the PDF $\rho_\infty$ of the stationary
distribution of~(\ref{eq:SDE-1d}) in the small noise limit. This
corresponds to the choice $O = \mathrm{id} : \RR \to \RR$ in our
formalism (with $d=d'=1$), and the stationary distribution is
available via $T \to \infty$.\\

From the Fokker-Planck equation
\begin{equation}
  \partial_t \rho(x,t) = \partial_x \left(V'(x) \rho(x,t) \right) + \eps \partial_{xx} \rho(x,t)\,,
\end{equation}
for the PDF of the process~(\ref{eq:SDE-1d}), the stationary distribution is known to be
\begin{equation}
  \rho_\infty(x) = \left[ \int_{-\infty}^\infty \dd x' \exp\left\{-\eps^{-1} V(x') \right\} \right]^{-1} \exp\left\{-\eps^{-1} V(x) \right\}\,,
\end{equation}
which, by applying Laplace's method on the prefactor in the limit $\eps \to 0$, becomes
\begin{equation}
  \label{eq:SDE-1d-result}
\rho_\infty(x) = (2 \pi \eps)^{-1/2} \left(V''(\bar{x}) \right)^{1/2} \exp \left\{- \eps^{-1} \left(V(x) - V(\bar{x}) \right) \right\}\,.
\end{equation}
We will reproduce this result, and in particular the prefactor, from our discussion in section \ref{sec:fluctuations} now. Note that the linear observable $O = \mathrm{id}$ does not leave any freedom at the right boundary of the time interval, so the BVP determinant $\det B$ from~(\ref{eq:final-result-including-bvp}) reduces to $1$. Similarly, the observable gradient reduces to 1, which means that~(\ref{eq:final-result-including-bvp}) and~(\ref{eq:final-result-wo-bvp}) are directly seen to coincide and yield
\begin{equation}
  \rho(x) = (2 \pi \eps)^{-1/2} Q(0)^{-1/2} \exp \left\{- \frac{1}{2} \int_{-T}^0 \dd t \; V'''(u_I(t)) p_I(t) Q(t) \right\} \exp \left\{- \eps^{-1} S_I(x) \right\}\,,
\end{equation}
where $Q$ solves the one-dimensional Riccati equation
\begin{equation}
 \dot{Q} = 2 - 2 V''(u_I) Q - V'''(u_I) p_I Q^2, \quad Q(-T) = 0\,.
\end{equation}
First, we compute the instanton trajectory: For the minimization problem
\begin{equation}
 u_I = \argmin_{\substack{u(-T)=\bar{x}\\u(0) = x}} S[u] = \argmin_{\substack{u(-T)=\bar{x}\\u(0) = x}} \frac{1}{4} \int_{-T}^0 \dd t\; (\dot{u} + V'(u))^2\,,
\end{equation}
the instanton equations that we obtain can be written as
\begin{equation}
  \begin{cases}
    \dot u_I + V'(u_I) = 2 p_I\\
    \dot p_I - V''(u_I) p_I = 0
  \end{cases}
\end{equation}
with boundary conditions $u_I(-T) = \bar{x}$, $u_I(0) = x$. Under our
assumptions on $V$ and in the limit $T \to \infty$, these equations
are solved by
\begin{equation}
\label{eq:gradient-inst}
 \dot{u}_I = V'(u_I) = p_I\,,
\end{equation}
such that the action at the instanton becomes
\begin{equation}
 S_I(x) = \frac{1}{4} \int_{-\infty}^0 \dd t\; (\dot{u}_I + V'(u_I))^2 = \int_{-\infty}^0 \dd t \; V'(u_I) \dot{u}_I  = V(x) - V(\bar{x})\,,
\end{equation}
which correctly reproduces the ${\cal O}\left(e^{\eps^{-1}} \right)$-term
in~(\ref{eq:SDE-1d-result}). Now, the easiest way to determine the
prefactor in this case is to go back to~(\ref{eq:gy-y-alpha-1})
because it is already linear in one dimension. In terms of $Y$ with $Q
= 2 Y / \dot{Y}$, the PDF can be written as
\begin{equation}
\label{eq:prefac-gradient-y}
 \rho(x) = (2 \pi \eps)^{-1/2} Y(0)^{-1/2} \exp \left\{\int_{-T}^0 \dd t \; V''(u_I(t)) \right\} \exp \left\{- \eps^{-1} S_I(x) \right\}\,,
\end{equation}
where $Y$ solves
\begin{equation}
\ddot{Y} - 2 (V''(u_I) \dot{Y} + V'''(u_I)p_I Y) = 0, \quad Y(-T) = 0, \; \dot{Y}(-T) = 2\,.
\end{equation}
Using~(\ref{eq:gradient-inst}), this becomes
\begin{equation}
\ddot{Y} = 2 \frac{\dd}{\dd t} \left(V''(u_I) Y \right)\,,
\end{equation}
which can directly be integrated to yield
\begin{equation}
\dot{Y} = 2 + 2 V''(u_I) Y,\quad Y(-T) = 0\,.
\end{equation}
Integrating once more, we obtain
\begin{equation}
Y(t) = 2 \exp\left\{2 \int_{-T}^t \dd s\; V''(u_I(s)) \right\} \int_{-T}^t \dd s\; \exp \left\{-2 \int_{-T}^s \dd \tau \; V''(u_I(\tau)) \right\}\,.
\end{equation}
The prefactor in~(\ref{eq:prefac-gradient-y}) can then be evaluated to
\begin{align}
\label{eq:prefac-gradient-y-simplified}
Y(0)^{-1/2} \exp \left\{\int_{-T}^0 \dd t \; V''(u_I(t)) \right\} = \left[2 \int_{-T}^0 \dd s \; \exp \left\{-2 \int_{-T}^s \dd \tau \; V''(u_I(\tau)) \right\} \right]^{-1/2}\,.
\end{align}
Since the instanton trajectory stays at the fixed point $\bar{x}$ for
an infinite amount of time in the limit $T \to \infty$, we approximate
\begin{equation}
  \int_{-T}^s \dd \tau \; V''(u_I(\tau)) \approx V''(\bar{x}) (s + T)\,,
\end{equation}
which, upon insertion in~(\ref{eq:prefac-gradient-y-simplified}), yields
\begin{align}
Y(0)^{-1/2} \exp \left\{\int_{-T}^0 \dd t \; V''(u_I(t)) \right\} \overset{T \to \infty}{\sim} &\left[2 \int_{-T}^0 \dd s \; \exp \left\{-2 V''(\bar{x}) (s + T) \right\} \right]^{-1/2} \nonumber\\
&= \left[\frac{1}{V''(\bar{x})} \left(1 - \exp \left\{-2 V''(\bar{x}) T \right\} \right) \right]^{-1/2} \nonumber\\
&\xrightarrow{T \to \infty} (V''(\bar{x}))^{1/2}\,.
\end{align}
This calculation correctly reproduces the prefactor
in~(\ref{eq:SDE-1d-result}). Note that in this case, the prefactor is
merely a normalization constant that does not depend on $x$, but we
were still able to determine this constant precisely with our
method. In contrast, in the numerical examples that we will consider
next, the prefactor does depend on the observable value where the PDF
is evaluated. First, however, we remark that we can also calculate the
prefactor for the one-dimensional gradient example using any $\alpha
\in [0,1]$, with the same result. Indeed, the general,
$\alpha$-dependent $Y$-equation~(\ref{eq:gy-y}) reduces to
\begin{align}
&\ddot{Y} + 2(1 - 2\alpha) V''(u_I) \dot{Y} - 2 \alpha V'''(u_I) V'(u_I) Y - 4 \alpha (1 - \alpha) \left(V''(u_I)\right)^2 Y = 0\,,
\end{align}
with initial conditions $Y(-T) = 0$, $\dot{Y}(-T) = 2$, and the naive approximations $V''(u_I) = V''(\bar{x})$ and $V'(u_I) = 0$ in the ODE give
\begin{equation}
  Y(t) = \frac{1}{V''(\bar{x})} \exp\left\{- 2 \alpha V''(\bar{x}) (t+T)\right\} \left(1 - \exp \left\{-2 V''(\bar{x}) (t+T) \right\} \right)\,,
\end{equation}
such that the $\alpha$-dependent terms in the prefactor
\begin{align}
Y(0)^{-1/2} \exp \bigg\{\alpha \underbrace{\int_{-T}^0 \dd t \; V''(u_I(t))}_{\approx V''(\bar{x}) T}\bigg\} \overset{T \to \infty}{\sim} V''(\bar{x})^{1/2}  \left[1 - \exp \left\{-2 V''(\bar{x}) T \right\} \right]^{-1/2}\,,
\end{align}
again tend to $V''(\bar{x})^{1/2}$, canceling out any
$\alpha$-dependence.

\subsection{Two-dimensional non-gradient system}

Here, we consider a two-dimensional, non-gradient SDE with a
one-dimensional observable as a second example, which we now treat
numerically. Motivated by future applications to stochastic PDEs, we
derive our example from the one-dimensional Burgers
equation~(\ref{eq:burgers-full-nondim}), but apart from this
motivation, the example has no physical significance and mainly serves
as a technical means to demonstrate the method at this point. We
transform the non-dimensionalized Burgers equation with periodic
boundary conditions on $[0,2 \pi]$ to Fourier space, which gives
\begin{equation}
\label{eq:burgers-fourier}
\frac{\dd}{\dd t} \hat{u}_k + \frac{i k}{4 \pi} \sum_{l \in \ZZ} \hat{u}_{k-l} \hat{u}_l + k^2 \hat{u}_k = \hat{\eta_k}\,,
\end{equation}
for $k \in \ZZ$ and $\hat u_k \in \CC$ the $k$-th Fourier
coefficient. Since the velocity field and the forcing are real, their
Fourier coefficients fulfill $\hat{u}_{-k} = \hat{u}^*_k$ and
$\hat{\eta}_{-k} = \hat{\eta}^*_k$. Now, by arbitrarily setting all
Fourier coefficients $\hat{u}_k$ with $|k| \geq 3$ to zero, we obtain
the two-dimensional complex SDE
\begin{equation}
\frac{\dd}{\dd t} \left(\begin{array}{c}
\hat{u}_1\\
\hat{u}_2
\end{array} \right)
+ \left(\begin{array}{c}
\hat{u}_1\\
4 \hat{u}_2
\end{array} \right)
+ \frac{i}{2 \pi} \left(\begin{array}{c}
\hat{u}_1^* \hat{u}_2\\
\hat{u}_1^2
\end{array} \right) = \left(\begin{array}{c}
\hat{\eta}_1\\
\hat{\eta}_2
\end{array} \right)\,.
\end{equation}
Note that this procedure can be interpreted as a Galerkin truncation
of the Burgers equation at the $k=2$ mode. In principle, apart from
numerical efficiency considerations, we could put the cutoff at any
number of modes.

A further reduction to a two-dimensional real system can by achieved
by considering only the antisymmetric parts of these two modes in real
space, which corresponds to keeping only the imaginary parts of their
Fourier coefficients. Dropping unnecessary constants for convenience,
we arrive at the two-dimensional real example
\begin{equation}
\label{eq:2-modes}
\frac{\dd}{\dd t}  \left(\begin{array}{c}
u_1\\
u_2
\end{array} \right)
+ \left(\begin{array}{c}
u_1\\
4 u_2
\end{array} \right)
+\left(\begin{array}{c}
u_1 u_2\\
- u_1^2
\end{array} \right) = \left(\begin{array}{c}
\eta_1\\
\eta_2
\end{array} \right), \quad \left< \eta(t) \eta^\top(t') \right> = \eps \mathrm{diag}(\chi_1, \chi_2) \delta(t-t')\,,
\end{equation}
where $u_1$ and $u_2$ are the imaginary parts of the Fourier
coefficients $\hat u_1$ and $\hat u_2$, respectively.  This system is
non-gradient, dissipative, and possesses only one stable fixed point
of the deterministic dynamics at $u_1=u_2=0$. The covariance matrix of
the forcing is chosen to be diagonal, as this will be the case for the
Fourier transform of a stationary forcing in real space. As a
one-dimensional observable, we approximate the gradient
\begin{equation}
\partial_x u(x=0,t=0) = - \frac{1}{\pi} \sum_{k=1}^\infty k \cdot \mathrm{Im}\left(\hat{u}_k(t=0)\right)\,,
\end{equation}
from~(\ref{eq:burgers-gradient-obs}) in terms of the two modes,
which yields, upon dropping the unnecessary constant,
\begin{equation}
\label{eq:2-modes-gradient}
 O(u) = -(u_1 + 2 u_2)\,,
\end{equation}
as the linear observable that we will consider in the following.\\

\begin{figure}
\centering
\includegraphics[width = \textwidth]{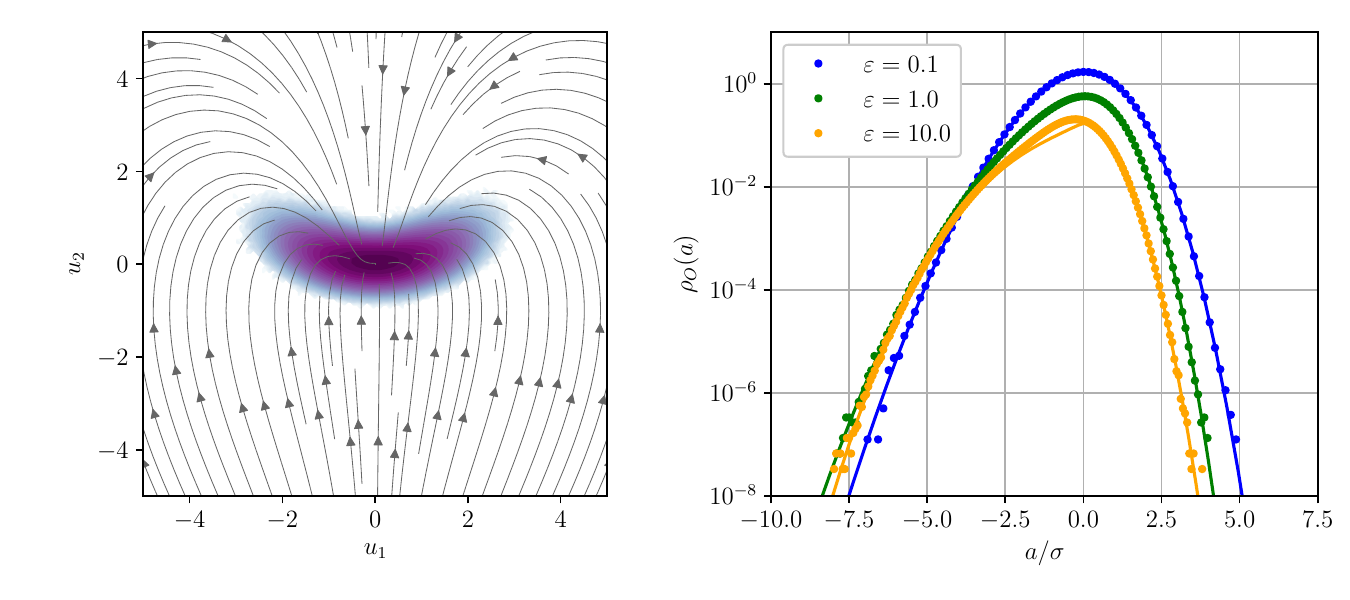}
\caption{Results of the Monte Carlo simulations of~(\ref{eq:2-modes})
  with $2 \cdot 10^8$ samples, and comparison to the instanton
  estimate including the prefactor $Z$. The left panel shows the
  distribution of $u(t=0)$ for $\eps = 1$. The right panel shows the
  PDFs $\rho_O$ for the
  observable~(\ref{eq:2-modes-gradient}) for different noise strengths
  $\eps$, scaled by their standard deviation $\sigma$. The Monte Carlo
  results are indicated by the data points, whereas the lines show the
  result of evaluating~(\ref{eq:final-result-wo-bvp}). We see that for
  the system at hand, there is an excellent agreement between the
  Monte Carlo results and the instanton estimate, even at high noise
  strengths, where slight deviations become visible only at $\eps =
  10$. A more precise comparison involving the prefactor itself can be
  found in Figure \ref{fig:z-results-2mb}.}
\label{fig:dns-pdfs}
\end{figure}

\begin{figure}
\centering
\includegraphics[width = \textwidth]{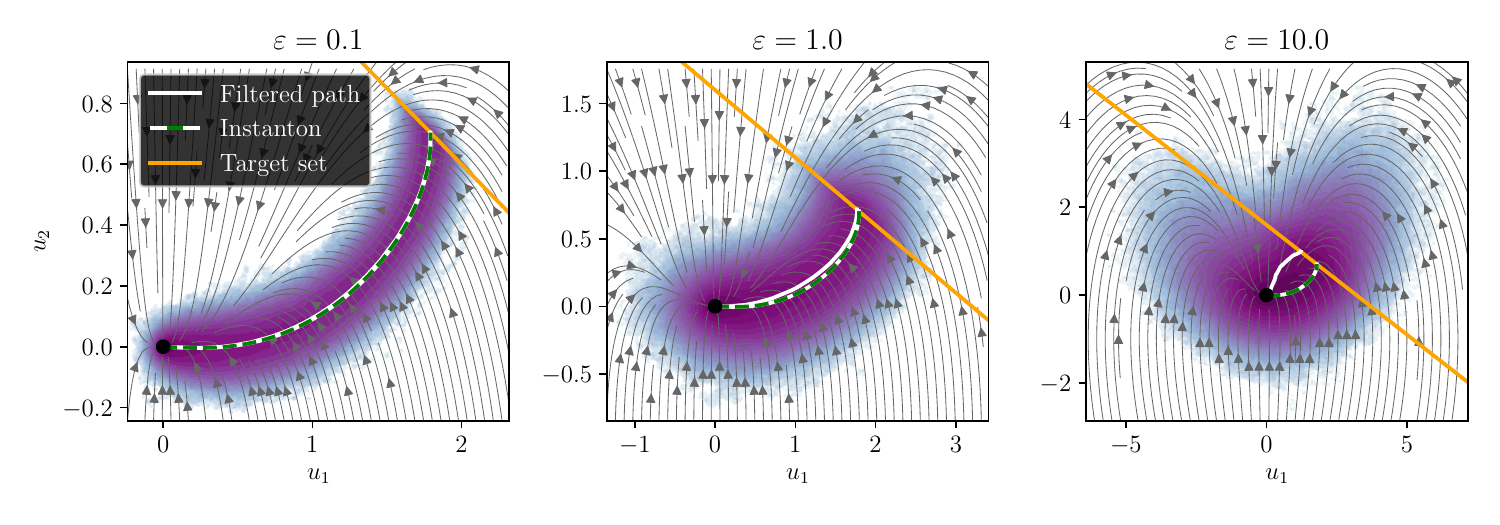}
\caption{Average paths from Monte Carlo simulations of~(\ref{eq:2-modes}),
conditioned on observable values~(\ref{eq:2-modes-gradient}) close to $a = -3.2$
(indicated by the orange line) at $t=0$ for different $\eps$, in comparison to the
instanton. The average was taken over $10^4$ samples, and the color plot shows
the two-dimensional histogram of the data. For a small noise amplitude, the instanton
and the filtered path agree well, and we can expect our quadratic approximation to
yield good results. At $\eps = 10$, the system is dominated by the noise for this
observable value and the instanton path does not provide a good approximation for
the filtered path.}
\label{fig:filter-2modes}
\end{figure}

For our numerical experiments, we took $T=1$, $\chi_k = k^{-2}$
and $u_0 = 0 \in \RR^2$ as the initial value, and considered three
different noise strengths $\eps \in \left\{0.1, 1, 10 \right\}$. For
each of the noise strengths, we performed $2 \cdot 10^8$ Monte Carlo
simulations of the SDE~(\ref{eq:2-modes}) in order to evaluate the PDF
$\rho_O$ at $t=0$. For these simulations, we used the stochastic Heun
scheme, together with an integrating factor for the linear,
dissipative terms, with a time step $\Delta t = 5 \cdot 10^{-4}$,
corresponding to $n = 2000$ discretization points in time. Figure
\ref{fig:dns-pdfs} shows the results of the Monte Carlo runs
for the PDF $\rho_O$, as well as the vector field $N$ for the
SDE~(\ref{eq:2-modes}) and the two-dimensional PDF of $u(t=0)$ itself for $\eps = 1$.
Furthermore, Figure \ref{fig:filter-2modes} shows the average path of
the process conditioned on hitting an observable value of $a = -3.2$
at $t=0$ for all three $\eps$, compared to the instanton path $u_I$
for that observable value\footnote{Note that for $\eps = 0.1$ and
$\eps = 1$, this observable value is already quite rare, so the ibis
method was used to determine the conditional expectation via
\begin{align*}
&\left<u(t_0)|O(u(0)) = a \right> = \frac{\left< u(t_0)\delta(O(u(0)) - a) \right>}{\left< \delta(O(u(0)) - a) \right>} \nonumber\\
&= u_I(t_0) + \sqrt{\eps} \frac{\left< \delta u(t_0) \delta(O(\delta u(0))) \exp \left\{-\frac{1}{2} \int_{-T}^0 \dd t \left(\delta u, (\nabla \nabla N(u_I), p_I)_d \delta u \right)_d \right\} \right>}{\left< \delta(O(\delta u(0)))\exp \left\{-\frac{1}{2} \int_{-T}^0 \dd t \left(\delta u, (\nabla \nabla N(u_I), p_I)_d \delta u \right)_d \right\}\right>}\,,
\end{align*}
for all $t_0 \in (-T,0)$, where $\delta u$ solves the nonlinear SDE~(\ref{eq:sde-ibis}).} \cite{grafke-grauer-schaefer:2013}.\\

Once the reference PDFs are obtained, in order to apply the methods
from section~\ref{sec:fluctuations}, we first need to compute the
instanton configurations over a range of relevant observable values
$a$. Note that our instanton approach with pre-factor estimate
necessitates only a single computation of the involved terms for all
noise strengths $\eps$, as the scaling in $\eps$ is given explicitly
in the PDF~(\ref{eq:final-result-including-bvp})
or~(\ref{eq:final-result-wo-bvp}). This is in contrast to Monte Carlo
simulations, which have to be performed for every noise strength
$\eps$ separately.

\begin{figure}
\centering
\includegraphics[width = \textwidth]{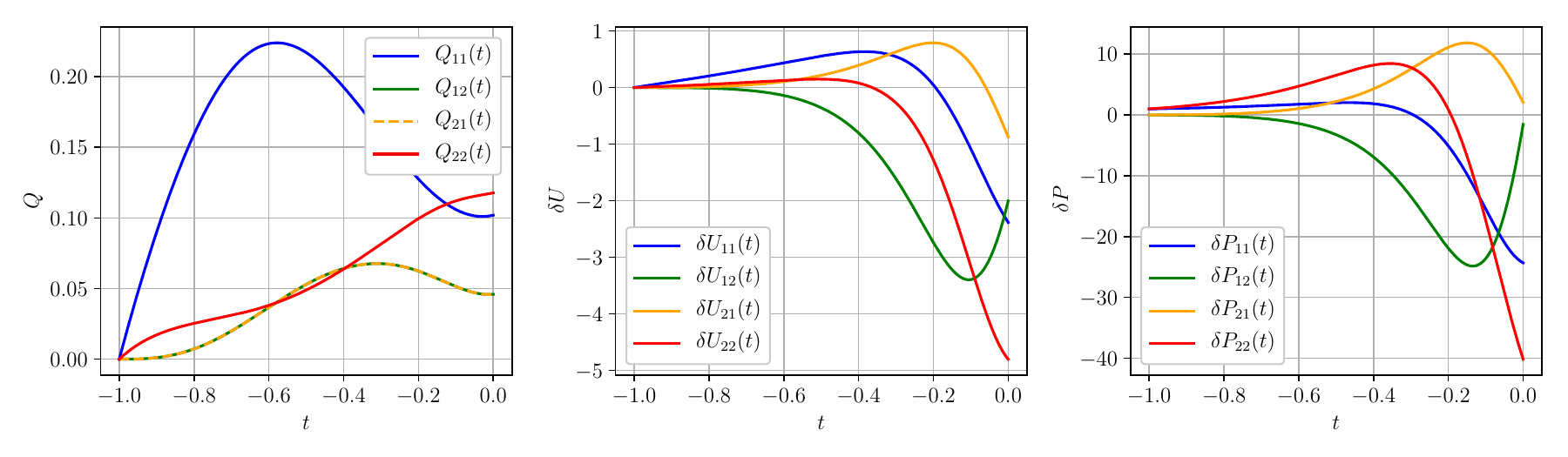}
\caption{The left panel shows the numerical solution of the Riccati
equation~(\ref{eq:riccati-w-final}) for the system~(\ref{eq:2-modes}) and
observable~(\ref{eq:2-modes-gradient}) at an observable value of $a = -11$.
The other two panels show the corresponding solution of the Radon transformed
linear equation~(\ref{eq:radon}) that corresponds to a classical Gel'fand-Yaglom equation.}
\label{fig:riccati-solution}
\end{figure}

For the numerical solution of the instanton optimization
problem~(\ref{eq:minimization-problem}), we incorporated the final
time constraint $O(u_I(t=0)) = a$ with a penalty approach and solved
the resulting unconstrained optimization problems with the L-BFGS
method \cite{nocedal-wright:2006}, which is an improvement over the
classical Chernykh-Stepanov \cite{chernykh-stepanov:2001} gradient
descent \cite{grafke-vanden-eijnden:2019}. The same parameters as
detailed above were used for the time discretization of the
optimization problem, and we checked that variations of the time
stepping scheme and time step size do not lead to appreciable
differences in the results. After these instanton trajectories, which
we computed for $350$ equally spaced values of $a \in [-20, 10]$, have
been calculated, we solve the Riccati
equation~(\ref{eq:riccati-w-final}) along each of these trajectories
in order to evaluate~(\ref{eq:final-result-wo-bvp}). Figure
\ref{fig:riccati-solution} shows a typical solution of the Riccati
equation for the system at hand. In order to evaluate the BVP
alternative numerically, we have to solve one BVP~(\ref{eq:bvp}) for
each $a$ since the system at hand is two-dimensional with a
one-dimensional observable.  The
observable~(\ref{eq:2-modes-gradient}) is linear, so its gradient does
not depend on $u_I(0)$, and the boundary value for which we need to
solve ~(\ref{eq:bvp}) is given by $\delta u^{(1)}_n = 5^{-1/2}
(2,-1)^\top$ for all $a$. In order to solve~(\ref{eq:bvp})
numerically, we use a simple shooting method. The PDF that we obtain
from~(\ref{eq:final-result-wo-bvp}), including the prefactor, is
directly compared to the respective PDFs obtained from direct Monte
Carlo simulation of~(\ref{eq:2-modes}) in Figure
\ref{fig:dns-pdfs}. We observe excellent agreement between the
instanton estimate and the actual PDFs, and now turn to a more
detailed analysis of the numerical results for the prefactor term.\\

\begin{figure}
\centering
\includegraphics[width = \textwidth]{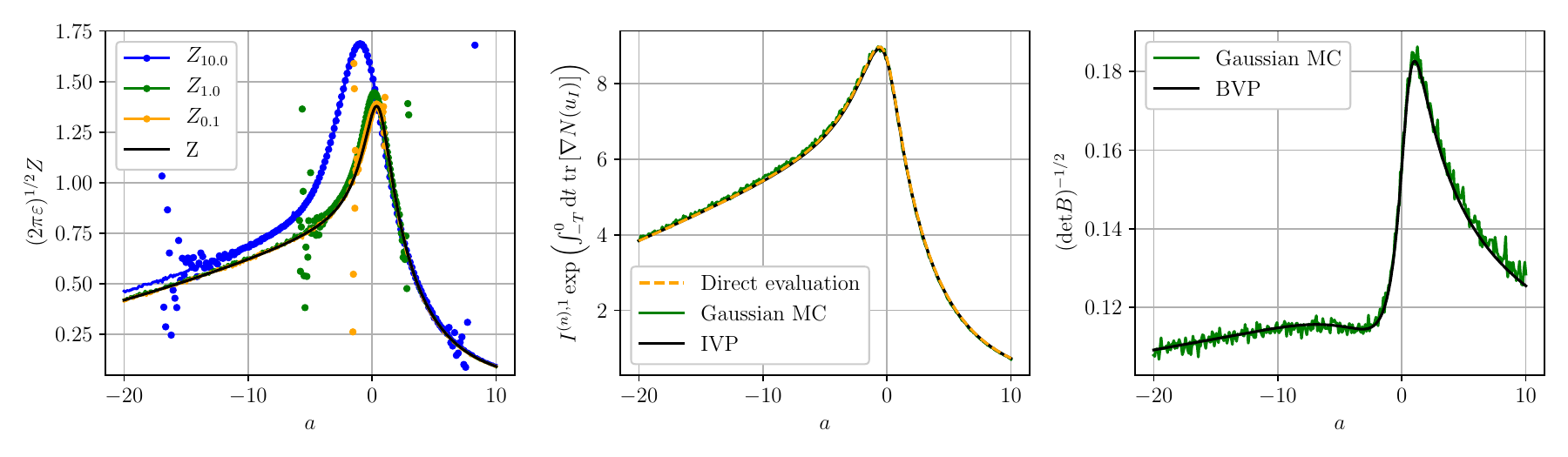}
\caption{Comparison of Monte Carlo results and our method for the prefactor of the
PDF of the linear observable~(\ref{eq:2-modes-gradient}) of the two-dimensional
SDE~(\ref{eq:2-modes}). The left panel shows the full prefactor $Z_\eps$
from~(\ref{eq:prefactor-general}) as obtained from direct Monte Carlo simulations
of~(\ref{eq:2-modes}) at different $\eps$, which is indicated by dots in the figure.
The lines of the same color show the results of the ibis method~(\ref{eq:sde-ibis})
for the same $\eps$'s in order to sample regions that are not accessible by the
direct simulations. For the latter, $2 \cdot 10^4$ samples were
taken for each observable value.
These Monte Carlo results are then compared to the quadratic prefactor $Z$ we obtain
from~(\ref{eq:final-result-wo-bvp}). The results of~(\ref{eq:final-result-including-bvp})
as well as Monte Carlo simulations of~(\ref{eq:sde-linear}) for the prefactor $Z$ coincide
with this (not shown). The second panel shows the specific contribution of the functional
determinant with Dirichlet 0 boundary conditions, together with the Jacobian, to
the total prefactor $Z$, which is either accessible by Monte Carlo simulations of the
observable $Z^{(0)}$ from ~(\ref{eq:obs-dirichlet-0}), by direct numerical computation
of the determinant of the matrix $H$ from~(\ref{eq:h-matrix-diagonal})
and~(\ref{eq:h-matrix-offdiagonal}), or, of course, by solving the
IVP~(\ref{eq:riccati-w-final}). Finally, the last panel to the right shows
the contribution of the fluctuations at the right time boundary, which we obtained
through the solution of one BVP~(\ref{eq:bvp}) for each $a$, and compare to
the quotient $Z/Z^{(0)}$ from Monte Carlo simulations.}
\label{fig:z-results-2mb}
\end{figure}

As already mentioned in sections \ref{subsec:continuum-overview} and
\ref{subsec:fluctuation-determinant-0bc}, there exist further
possibilities to individually access the prefactor, the BVP
determinant~(\ref{eq:det-b}) and the functional determinant with
Dirichlet 0 boundary conditions $\det H$ from~(\ref{eq:int-dirichlet})
numerically, in order to be able to compare these individual terms to
the expressions which we derived. First, the full prefactor
$Z_\eps$ for $\eps > 0$, defined
in~(\ref{eq:prefactor-general}), is numerically available either by the
results of the direct numerical simulations of~(\ref{eq:SDE}) that we
performed, or, in observable ranges that are not sufficiently sampled
for a specific $\eps$, by the ibis approach~(\ref{eq:sde-ibis}).  For
the quadratic prefactor $Z$, we can then either
simulate~(\ref{eq:sde-linear}) for a Monte Carlo approach, solve the
BVP~(\ref{eq:bvp}) and the Riccati equation~(\ref{eq:riccati-w-final})
and evaluate~(\ref{eq:final-result-including-bvp}), or only solve the
Riccati equation and compute $Z$ from~(\ref{eq:final-result-wo-bvp}).
The results of these different approaches for the 2-mode
system~(\ref{eq:2-modes}) and the ``velocity gradient''
observable~(\ref{eq:2-modes-gradient}) are shown in Figure
\ref{fig:z-results-2mb}. Finally, by simulating~(\ref{eq:sde-linear})
with the observable
\begin{align}
\label{eq:obs-dirichlet-0}
Z^{(0)} = \left< \delta(\delta u(0)) \exp \left\{-\frac{1}{2} \int_{-T}^0 \dd t \left(\delta u, (\nabla \nabla N(u_I), p_I)_d \delta u \right)_d \right\}\right>\,,
\end{align}
we can evaluate the integral~(\ref{eq:int-dirichlet}) with Dirichlet
0 boundary conditions (multiplied by the
Jacobian $\exp \{ \alpha \int_{-T}^0 \dd t \; \trace[\nabla N(u_I)]\}$),
in order to compare it to the Gel'fand-Yaglom result~(\ref{eq:gy-result}), as well as a
direct numerical computation of the determinant of the $(n-1)d \times (n-1)d$ matrix
$H$ as defined by~(\ref{eq:h-matrix-diagonal}) and~(\ref{eq:h-matrix-offdiagonal}).
The quotient of $Z$ and $Z^{(0)}$, as determined from Monte Carlo simulations,
should then be given precisely by the BVP determinant $(\det B)^{-1/2}$,
which is also shown in Figure \ref{fig:z-results-2mb}.\\

\subsection{Preliminary results for the full Burgers equation}

\begin{figure}
\centering
\includegraphics[width = \textwidth]{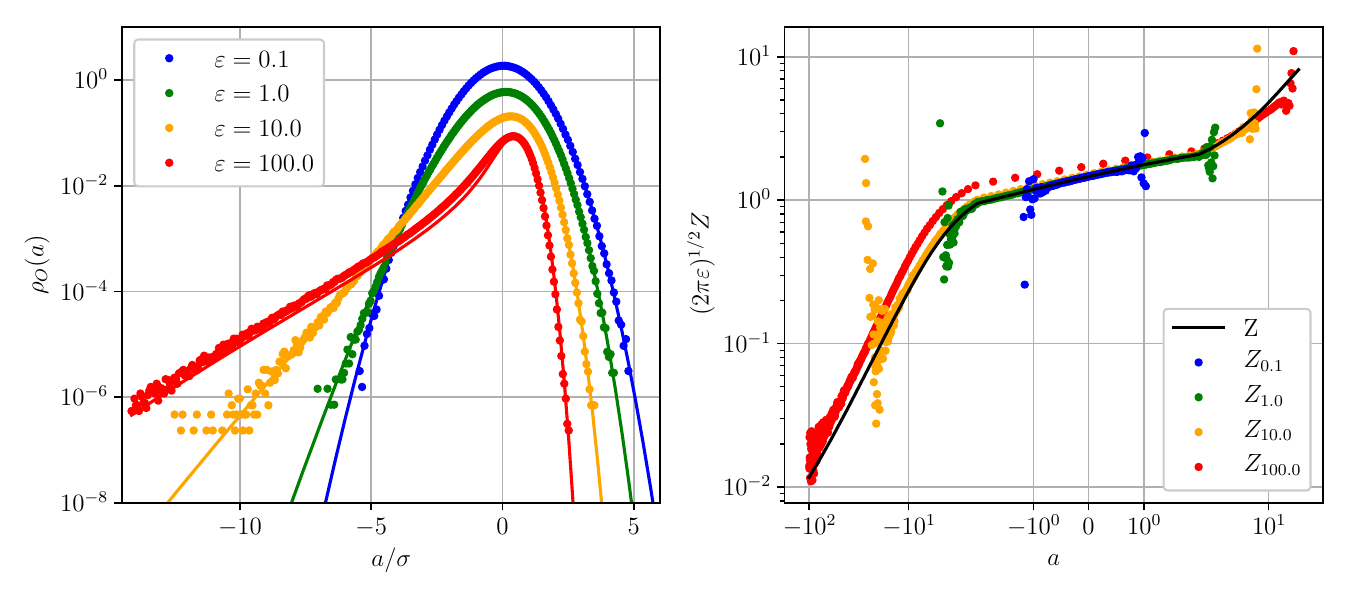}
\caption{Numerical results for the Burgers equation~(\ref{eq:burgers-full-nondim}).
The left panel shows the PDFs of the gradient observable $O(u) = \partial_x u(x=0, t=0)$
for different noise strengths $\eps$, scaled by their respective standard
deviation~$\sigma$. Using the normalization
from \cite{grafke-grauer-schaefer-etal:2015}, the Reynolds number
corresponding to these noise strengths is given by $\mathrm{Re} = \eps^{1/3}$.
For each $\eps$, we performed $5 \cdot 10^5$ Monte Carlo simulations
with a spatial resolution $n_x = 64$ and $n = 1000$ Heun time steps
(with integrating factor for the dissipative term) for $T=1$.
The results of these Monte Carlo simulations are indicated by the dots
in the left figure, whereas the lines of the same color show the result of
evaluating~(\ref{eq:final-result-wo-bvp}). Note that, as in
Figure \ref{fig:dns-pdfs}, deviations of the Monte Carlo results
from~(\ref{eq:final-result-wo-bvp}) only become visible at large $\eps$,
in this case at $\eps = 100$. The right panel shows the full
prefactor $Z_\eps$ from~(\ref{eq:prefactor-general}), as obtained
from these Monte Carlo simulations, in comparison to the quadratic
prefactor from~(\ref{eq:final-result-wo-bvp}) on a log-log scale.}
\label{fig:burgers-results}
\end{figure}

Here, we show preliminary results for the prefactor calculation method from
section~\ref{sec:fluctuations}, applied to the full Burgers
equation~(\ref{eq:burgers-full-nondim}) on $[0, 2 \pi]$ with periodic boundary
conditions at a relatively small spatial resolution $n_x = 64$ (i.e.\ we
have $d = 64$ for this example in the notation of the previous sections).
The specific resolution that we used here was chosen arbitrarily; extending
the prefactor calculation method to higher spatial resolution poses no
conceptual or numerical problems, at least for one-dimensional SPDEs.
In this example, we choose the Mexican hat function
\begin{equation}
\chi(x) = - \partial_{xx} \left( \exp\left\{- \frac{x^2}{2} \right\} \right) = \left(1 - x^2 \right) \exp\left\{- \frac{x^2}{2} \right\}\,,
\end{equation}
for the large-scale spatial correlation function of the noise, and perform
pseudo-spectral Monte Carlo simulations of the Burgers
equation~(\ref{eq:burgers-full-nondim}) at different noise strengths, or,
equivalently, at different Reynolds numbers, in order to evaluate the
PDF of the gradient of the velocity field~(\ref{eq:burgers-gradient-obs}).
The results of these simulations, as well as a comparison to the results of the
corresponding instanton and prefactor computations, can be found in
Figure \ref{fig:burgers-results}. Note the excellent agreement between
the Monte Carlo results and the instanton estimate, both for the full PDF
and for the prefactor, even at relatively large $\eps$. As in the previous
example, the instanton configurations were computed from a variant of the
classical Chernykh-Stepanov algorithm over a range of relevant observable
values $a$, and then, for each $a$, the Riccati equation~(\ref{eq:riccati-w-final})
was integrated in order to evaluate~(\ref{eq:final-result-wo-bvp}). Due to
the fact that we approximate the partial differential
equation~(\ref{eq:burgers-full-nondim}) in this example, the inner products
$(\cdot, \cdot)_d$ in~(\ref{eq:riccati-w-final}) and~(\ref{eq:final-result-wo-bvp})
were modified by an additional factor $\Delta x = 2 \pi / n_x$ in this case.
Concretely, this means that for $Q: [-T, 0] \to \RR^{n_x \times n_x}$, we integrate
\begin{equation}
\dot{Q} = \chi - \left[ \nabla N(u_I) Q + \left(\nabla N(u_I) Q\right)^\top  \right] - \Delta x \cdot Q (\nabla \nabla N(u_I), p_I)_d Q,
\end{equation}
where $\chi \in \RR^{n_x \times n_x}$ is a Toeplitz matrix
with $\chi_{kl} = \chi((k - l) \cdot \Delta x )$, and $\nabla N(u_I) Q$
as well as $(\nabla \nabla N(u_I), p_I)_d Q$ are evaluated column-wise by
means of Fast Fourier Transforms. The prefactor is then evaluated as
\begin{align}
 Z &= (2 \pi \eps)^{-1/2} \exp \left\{- \frac{1}{2} \int_{-T}^0 \dd t \; \Delta x \; \trace \left[ (\nabla \nabla N(u_I(t)), p_I(t))_d Q(t) \right] \right\} \times \nonumber\\
  & \quad \times
  \left[\nabla O(u_I(0))Q(0)\nabla O(u_I(0))^\top\right]^{-1/2}\,,
\end{align}
where, as $n_x \to \infty$, the last factor amounts to an
evaluation of $\partial_{xy}Q(x = 0, y = 0, t=0)$ for the linear
observable~(\ref{eq:burgers-gradient-obs}).

\section{Discussion and Outlook}
\label{sec:conclusion}
Here, we briefly summarize and discuss the results of this paper
and provide an outlook on further related questions.
Even though the instanton method is well established in the literature in
order to estimate observable PDFs of SDEs in a suitable large deviation limit,
general procedures to obtain sharper estimates for these PDFs by including the full
prefactor $Z$ at leading order have not been investigated systematically in this context
up until now. For the case of Langevin-type SDEs with additive white-in-time Gaussian
noise and unique instanton solutions, we fill this gap with the proposed
method. In principle, apart from the unwieldy discretized expressions
that we encountered in the derivation of our
main result, our approach consists of a straightforward and conceptually simple
evaluation of the Gaussian path integral that is obtained by expanding the action
to second order around the instanton trajectory. Using a variant of the traditional
Gel'fand-Yaglom approach to calculate such path integrals, we were able to reduce
the path integral evaluation to the solution of a matrix Riccati differential
equation as an initial value problem, which turned out to be possible even for
the boundary conditions $\nabla O(u_I(0)) \delta u(0) = 0$ on the right boundary
of the time interval that we encountered in the specific application of calculating
low-dimensional observable PDFs. Numerically, computing the prefactor $Z(a)$ at
an observable value $a \in \RR^{d'}$ with the proposed method thus amounts to the
solution of a single initial value problem of size $d \times d$ in addition
to the computation of the instanton itself, which is easily possible for moderately
large system dimensions (stemming from the discretization of one-dimensional SPDEs)
and in fact much cheaper than the iterative computation of the instanton trajectory.
We then proceeded to apply the prefactor calculation method to examples of one-dimensional
and two-dimensional SDEs, where the former was treated analytically, whereas, for
the latter, we showed detailed numerical results to test the predictions
of our prefactor calculation method against Monte Carlo results and direct numerical evaluations of the fluctuation matrix determinant.
Afterwards, we showed first results for the important
example of the velocity gradient PDF in one-dimensional Burgers turbulence,
which already appear quite promising and will be expanded upon in future studies.

In this regard, one of the ultimate questions is what maximum Reynolds numbers can be
achieved, and whether this is a possible way to understand
intermittency in turbulence. A related question is whether it is
possible to recover the high Reynolds number $7/2$ inviscid scaling
of the gradient PDF in Burgers 
turbulence~\cite{e-vanden-eijnden:1999,bec-khanin:2007} using this
approach. Here, on the technical side, it is not clear whether the direct 
solution of the
matrix Riccati equation (\ref{eq:riccati-w-final}) or the solution of
the Radon-transformed linearized system (\ref{eq:radon}) is more
advantageous. The linearized system (\ref{eq:radon}) would have the
enormous advantage of being ideally suited for parallel calculations,
but difficulties may arise due to the appearance of the backward heat
equation hidden in the term $\nabla N^\top$. It should be mentioned,
however, that the matrix Riccati equation (\ref{eq:riccati-w-final})
is also amenable to a massive parallel approach~\cite{benner-mena:2004,granat-kagstrom-kressner:2008} or
tensor network techniques~\cite{breiten-dolgov-stoll:2020}. The
ultimate challenge would be the application of our approach to the
full three-dimensional Navier-Stokes equations.

\begin{acknowledgments}
  The authors thank the anonymous referees for their comments
  and for pointing out important literature.
  The authors thank Sandra May for helpful discussions regarding the
  constrained instanton optimization problem. TG thanks Tobias Sch\"afer
  and Eric Vanden-Eijnden for helpful discussions and acknowledges the
  support received from the EPSRC projects EP/T011866/1 and
  EP/V013319/1.
\end{acknowledgments}

\bibliography{bib}

\end{document}